# Future Web Growth and its Consequences for Web Search Architectures

**Andrew Trotman**
Department of Computer Science
University of Otago
Dunedin, New Zealand

**Jinglan Zhang**
Science and Engineering Faculty
Queensland University of Technology
Brisbane, Queensland, Australia

**Abstract**

**Introduction**. Before embarking on the design of any computer system it is first necessary to assess the magnitude of the problem. In the case of a web search engine this assessment amounts to determining the current size of the web, the growth rate of the web, and the quantity of computing resource necessary to search it, and projecting the historical growth of this into the future.
**Method**. The over 20 year history of the web makes it possible to make short-term projections on future growth. The longer history of hard disk drives (and smart phone memory card) makes it possible to make short-term hardware projections.
**Analysis**. Historical data on Internet uptake and hardware growth is extrapolated.
**Results**. It is predicted that within a decade the storage capacity of a single hard drive will exceed the size of the index of the web at that time. Within another decade it will be possible to store the entire searchable text on the same hard drive. Within another decade the entire searchable web (including images) will also fit.
**Conclusion**. This result raises questions about the future architecture of search engines. Several new models are proposed. In one model the user's computer is an active part of the distributed search architecture. They search a pre-loaded snapshot (back-file) of the web on their local device which frees up the online data centre for searching just the difference between the snapshot and the current time. Advantageously this also makes it possible to search when the user is disconnected from the Internet. In another model all changes to all files are broadcast to all users (forming a star-like network) and no data centre is needed.

## 1. Introduction

Since its inception in the early 1990s the World Wide Web (the web) has become an indispensable information resource for many people. In 2010 there were an estimated over 2 billion users of the web and about 770 million hosts (Central Intelligence Agency (US) 2011). The web has fundamentally changed the way we communicate; and continues to do so.

Web browsers on mobile phones are increasingly used to access web content. According to the International Telecommunications Union[1], in 2010 there were (globally) 78 mobile phone subscriptions per 100 head of population but only 17.2 land-line subscriptions per 100. Mobile phones have become an essential communication technology for people in the poorest countries because land lines do not exist and it is cheaper to install a cell than to string land lines. These countries have large populations but low Internet coverage, but this is changing.

Eventually Internet penetration will saturate. So too will the world population. This leads us to believe that the growth rate of *human-generated* information on the web will stabilize.

This stable point is of particular interest because it bounds the number of uses of an online search engine, and bounds the growth rate of information in that search engine. Factoring in the increase in computing power and hard disk capacity raises the research question:

*Will the search problem be harder or easier next year?*

It is essential to answer this question and the related question

*Is the trend towards harder or easier?*

before embarking on the design of a new web search engine.

Herein these questions are examined by collating, where possible, the raw or summarized statistics published by reputable data providers. As an example, to determine the world population the data from the United Nations is used. Because data is sourced directly from providers it has not undergone scientific scrutiny, however it is more recent than older data already discussed in the literature.

This data is used to first estimate the world population and to project that into the future. World population provides a cap of the maximum possible number of Internet users. Then the history of the number of web pages indexed by the search engines, and the size of those pages, are examined. The indexable pages are of particular interest because they are (intended to be) spam and duplicate free and so a good measure of the sum of human knowledge. This data along with the Internet user base at the time lead to projections of the number of documents and their size that can be expected to exist in the future.

The size of the inverted index of the web is estimated by examining multiple search engine indexes over several collections. From this, and the projected size of the web, an estimate of the index size for the whole web and its future growth are made. At this point the size of the web and the size of its index (now and into the future) are known.

Next the history of hard disk drive size and of smart phone memory card size is plotted and projected into the future. It is shown that hard disk drive capacity grows at a faster rate than the web accumulated content – the two graphs cross. If the future projections are accurate then within 10 years the web index will be smaller than the size of a single hard disk drive. Within another 20 years it will be possible to store the entire indexable web and the index on a single hard drive.

---

[1] http://www.itu.int/ITU-D/ict/statistics/

CPU throughput and network bandwidth are then discussed. Then finally the design consequences of the hardware and data projections are examined and new web search engine architectures are proposed. These architectures involve pre-loading pages onto each hard drive on manufacture (much as the operating system is) and using the client computer as part of the distributed search architecture. Date based sharding is suggested as a method to make it possible to search jointly across the user's snapshot and a centralized index of the updates since the user's snapshot. Update mechanisms are discussed.

Eventually this architecture will be possible on smart phones. When this happens it will be possible to search and retrieve most of the web even if the user is in a place that has no Internet connectivity (such as in space or under water).

The contributions of this paper include:
1. The collection of historical data from multiple sources showing the world population and Internet uptake. From this the future growth of the Internet is predicted.
2. The collection of historical data from multiple sources showing hard disk drive and SD memory card growth. From this the future growth of these cards is predicted.
3. The demonstration that hard drives are growing faster than the web and the prediction of when drives will be larger than the web.
4. New distributed search engine architectures that include the client machine.
5. The consequences of this new architecture, including the hypothesis that it will be smaller and more energy efficient than existing data centres.

Herein only the size of the human generated web is of interest. In particular, only the publically visible web, i.e., the part of the web that can be crawled and indexed by a search engine, is considered. The part of the web that is hidden from conventional search engines, such as password-protected or access-fee-involved databases is not considered. The Internet of things and other machine generated raw data, spam, and mirrors is also not considered. However, the human generated web extends beyond static pages; for example, the detailed information on books found on amazon.com is undoubtedly generated from a database maintained by humans. Such collections are indexed by web search engines and so are of interest and are included (however mirrors are not).

The remainder of this paper is organized as follows: The next section presents related work. Section 3 quantitatively investigates Internet growth by estimating the number of users, the kinds of pages they create, and the creation rate. Section 4 examines the history of storage device capacity and predicts when it will be possible to store the entire indexable web on a single storage device. Section 5 discusses the possible error in the estimates made in section 3 and 4. Sections 6 and 7 discuss CPU and networking respectively. Section 8 proposes the new search architectures. Finally section 9 concludes the paper.

## 2. Related Work

There are two threads of related work: research on the growth of the web; and work on the design of search engines. Both are discussed briefly in this section and throughout the remainder of this paper.

## 2.1. The Size of the Web

Research measuring the size of the Web and its traffic has contributed to the understanding of the impact of the Web on society as well as the effect of Web traffic on Internet resources.

Many have measured the size of the web by counting the number of web pages. Lawrence & Giles (Lawrence and Giles 1999) sampled the publicly indexable web and measured the number of servers, number of web pages, web page size, and the size of resources such as images within a web page. The estimated number of servers at February 1999 was 2.8 million. Each web server hosted an average of 289 web pages. The total number of indexable web pages was about 800 million. The mean size of a page was 18.7 KB (median 3.9 KB), or 7.3 KB (median 0.98 KB) after reducing the pages to text (removing HTML tags, comments, extra white space, etc.). Their work showed that the number of pages indexed by the popular search engines at that time was very low (maximum 16%) compared to the number of web pages estimated to exist.

Bharat & Broder (Bharat and Broder 1998) examined the number of Web pages indexed by the largest search engines (including Hotbot, Altavista, Excite, and Infoseek) and estimated about 200 million pages. They also discovered that the overlap between the search engines was low (1.4%). Seven years later Guilli & Signorini (Gulli and Signorini 2005) also estimated the relative size and overlap of the largest web search engines and updated the estimate to 11.5 billion pages. Using similar techniques de Kunder (de Kunder 2006) developed a live system that daily measures the size of the web by estimating the number of web pages indexed by the major search engine providers. The system has been running since 2006 and provides historic counts. As of June 2011 the estimated number of indexed web pages is between 17 billion and 46 billion depending on the estimation technique.

Government organizations such as the ITU (International Telecommunication Union 2011) as well as several commercial organizations (e.g. comScore[2]) have actively monitored the growth of the web. ITU measures the Internet size as the number of connected computers and the number of users, and they provide yearly reports on this. The estimate for 2010 is 2,084 million users, 940 million mobile broad band subscriptions, 555 million fixed broadband subscriptions, and 1,197 million fixed telephone lines. The ITU also reports 5,282 million mobile cellular subscriptions, many times the number of fixed telephone lines.

DomainTools[3] believe they have the most comprehensive collection of domain name ownership records. They measure the growth of the web by monitoring the daily changes in the top level domains including: .com, .net, .org, .info, .biz, and .us. In June 2011, the total number of registered names in these domains was 130,207,722[4].

While the number of users, hosts and domains (sites), and pages are good measures of Internet size, they do not measure usage. Coffman & Odlyzko (Coffman and Odlyzko 1999) measure the size and growth rate of the Internet in terabytes per month on public Internet and private line networks carrying voice and data traffic in the US. Their study focuses on network bandwidth capacity and the traffic they carry. They estimate the growth rate in public Internet traffic to be about 100% per year (at that time).

---

[2] www.comscore.com
[3] www.domaintools.com
[4] www.dailychanges.com

comScore[5] measures Internet traffic growth by counting the number of worldwide searchers and queries. They report that in July 2009 there were more than 113 billion searches conducted by people age 15 and over coming from home and work locations. They estimate this being equivalent to 103.3 searches per user conducted over 11 days of the month.

Even though the same metrics (such as number of web pages) are used by many of these studies, the results vary depending on the location, scale, and date of the work. It is generally acknowledged that it is difficult to measure the size of the Internet and the web.

## 2.2. Distributed Search

Several different distributed search engine architectures have been proposed and are in use.

The large commercial search engines (such as Google, Yahoo!, and Bing) use multiple data centres strategically placed around the world. Each data centre consists of thousands of identical commodity computers and costs hundreds of millions of dollars to build (Strassmann 2005). These commercial search engine companies embrace the distributed computing (or cloud) paradigm with data and computation distributed among multiple servers in multiple locations (Ghemawat, Gobioff et al. 2003; Isard, Budiu et al. 2007; Shvachko, Kuang et al. 2010).

The search engine itself consists of at least three parts: a web crawler for collecting information, a file system or database to store information, and an indexing / searching system to find documents relevant to a query. Accordingly, they have at least three types of cloud (Gu and Grossman 2008): a storage cloud to provide file-based services; a data cloud to provide data management services; and a compute cloud to provide computation services. Several models for this three-cloud approach are in use.

Amazon[6] includes the S3 storage cloud, SimpleDB data cloud, and EC2 compute cloud. The open source Hadoop project[7] includes: the Hadoop Distributed File System, HBase and Hadoop's MapReduce. The University of Illinois includes the Sector storage cloud and Sphere compute clouds (Gu and Grossman 2008). Multiple clouds of the same organization are often stacked together to form an overall computing infrastructure.

### 2.2.1 Google

A substantial body of work exist on nearly every aspect of Google, and it makes a good case study of how a large scale commercial search engine works.

Until recently Google included: the Google File System (Ghemawat, Gobioff et al. 2003); the BigTable database system built on the Google File System and designed to scale to petabytes of data on thousands of machines (Chang, Dean et al. 2006); and the MapReduce infrastructure (Dean and Ghemawat 2004), a framework for large-scale parallel computation on BigTable.

Google's architecture consisted of data gathering servers that crawled the web and downloaded pages for storage in document databases on document servers. These were then indexed later (Brin and Page 1998).

As the capacity of a single machine was too small to store the entire collection, the document collection was broken into small subsets of documents called shards (Barroso, Dean et al. 2003).

---
[5] www.comscore.com/Press_Events/Press_Releases/2009/8/Global_Search_Market_Draws_More_than_100_Billion_Searches_per_Month

[6] http://aws.amazon.com

[7] http://hadoop.apache.org

Each shard was then replicated at least three times for reliability and load balancing (Strassmann 2005). The shards were indexed and the index stored on index servers. Each index server contained a set of index shards. The index servers were also replicated.

Web servers coordinated the execution of queries sent by users by choosing which replicas to use to resolve the query and to produce the results lists (Barroso, Dean et al. 2003).

Finally, Google also included several other auxiliary servers to perform tasks such as spelling correction and advertisement serving.

Sharding is a particularly appealing architecture as it is shared-nothing. A shard has (and needs) no knowledge of the complete collection to operate. The design scales well to both an increasing number of documents (by adding shards) and to an increasing number of users (by adding replicas). Random and uniform distribution of documents to shards is simple and has been shown to be resilient to failure (Barroso, Dean et al. 2003).

An early Google innovation was to keep the entire index in memory. This removed any latency due to the mechanics of a hard disk drive (seeking and reading). This, in turn, increased throughput and decreased latency in queries (Dean 2009).

In June 2010 Google introduced their next-generation indexing system called "Caffeine". It continuously crawls and updates the search index (Grime 2010). Caffeine[8] replaces MapReduce but uses BigTable. More new storage architectures and indexing paradigms continue to be investigated as user requirements and hardware evolve (Fikes 2010).

### 2.2.2 Yahoo!

Baeza-Yates et al. suggest an alternative distributed model which uses a large number of small data centres located close to end users (Baeza-Yates, Gionis et al. 2009; Cambazoglu, Plachouras et al. 2009; Sarigiannis, Plachouras et al. 2009; Baeza-Yates 2010). This model has a geographically distributed search architecture with distributed query processing.

Their architecture has a distributed but localized crawling and indexing model. Unlike a centralized search engine, (where the index is stored in a single large data centre), their index is partitioned into small non-overlapping parts and distributed over multiple geographically distant data centres. Each data centre only crawls web pages stored on servers geographically close to it. Consequently, a data centre in France crawls and indexes documents in Europe. To this index they add a replica of the most popular parts of the whole web.

This model includes distributed query processing. User queries are mapped to local data centres according to the users' geographical location (e.g. queries from Europe being mapped to the data centre in France). If this data centre can process the query locally then it returns results without contacting other centres; otherwise the local site forwards the query to data centres likely to contain relevant results. They call this selective query forwarding.

The benefits of their model come from the proximity of the search engine to the Web data and its users. Crawling is efficient because Internet connections are fast. Local searching results in lower network utilization, shorter response time, and reduced communication cost. Most importantly it results in services (index content) tailored to local culture and language. The drawback is the potential problems associated with cross-data centre communications. Cambazoglu & Baeza-Yates (Cambazoglu, Baeza-Yates et al. 2011) discuss difficulties and unsolved problems with the centralised model and the Yahoo! distributed model.

---

[8] The Register, Google search index splits with MapReduce, http://www.theregister.co.uk/2010/09/09/google_caffeine_explained/

*2.3. High Performance Computing*

Models for high performance computing have evolved from the 1970s centralized main frame super-computer paradigm, to the 1990s Network Of Workstations (NOW) and Pile Of PCs (POPC) distributed computing, to the 2010s centralized-distributed compute cloud.

The 1970s super-computer model assumed that high performance processors were expensive and scarce hence they needed to be shared. Data was stored on inexpensive storage servers and moved to the powerful processor when the processor became available. When the computation finished, the results were moved back to storage server. The model separated compute servers from data servers (which were often stored geographically elsewhere). It was essentially a distributed data, sequential processing model and as such the weakness was an Input / Output bottleneck.

The 1990s NOW data centre model co-located data and computation within one data centre (and server) when possible. It made data movement parallel among multiple compute servers therefore overcoming the I/O bottleneck of the supercomputing model. The data centre model is a distributed data, parallel computation model. It is flexible and powerful but only suited to certain tasks – the so-called embarrassingly parallel tasks. Searching large quantities of text to determine the presence or absence of a set of keywords is such a task.

The 2010s cloud model provides otherwise awkward to manage services on top of the data centre model. These services include distributed replication, reliability, and the ability to scale to increased workload on demand.

Problematically, all these models assume the user is online. This has been a reasonable assumption because the resources necessary to index and search the web are only available to a tiny number of large organizations, but in the future, as we show, this amount of computing power will be available *en masse*.

With the continued increase in capacity of hard disk drives and the increased throughout of CPUs, powerful, lightweight, and small scale handheld devices will become available. If Moore's law continues to hold then these devices will (eventually) surpass the power of today's clouds. If this is the case then it will be possible to search today's web on those future devices. If this is the case then Internet connectivity no longer need be an obstacle to information access; as the web and its index could be pre-loaded into this device. In that future people in the most remote places of the poorest countries will not be hindered by lack of land-line connectivity in order to access the collective knowledge of mankind.

Today's web will be of little use to those future users if it takes hundreds of years to realize this kind of hardware; it is the web of their own time they will want. To realize that hardware will take longer, by which time the web will be larger, and so on[9]. This leads to the question of which is growing faster, the hardware or the quantity of textual information on the Internet.

To extrapolate to this future it is necessary to study the past. In the next section an estimate of the size of the web is made. The quantity of text is examined. The growth rate is examined. And finally the size of the inverted index necessary to search it is made. But first the size of the user base is estimated – and to determine that the world population size must be known.

---

[9] Indeed, by Aristotle's Achilles and the Tortoise paradox we can never catch up; and this is the topic of investigation.

## 3. Quantitative Internet Growth

Web content is created by users and there is a maximum rate at which those users can create that information. This section first estimates the number of users, then the kinds of pages they create – and at what rate.

### 3.1. World Population Growth

Good estimates of the world population size and growth rate come from the United Nations. The 2008 revision of the official population estimates (United Nations 2011) has been used for many years. However, 2008 is a misnomer.

To predict the world population the United Nations collected the census and register information for each country, factored in data for fertility, mortality and migration, and predicted as accurately as possible the world population as of 1 July 2010.

The reported current world population is an estimate, and the future population estimates the United Nations give are speculative. Because of this they provide several models of fertility, mortality, migration, and so on. Over a long period of time variations on these parameters have a substantial effect. For the purpose of simplicity the medium variant model is used herein.

In the medium model the birth rate of all women over all countries is predicted to reach to 1.85 children by 2050. The trajectory of the birth rate in currently high birth rate countries is assumed to follow the world trend of countries that fell between 1950 and 2010; but may not reach 1.85 by 2050. In currently low birth rate counties the birth rate is assumed to rise to 1.85 children by 2050; but may not reach 1.85 by 2050. In both cases, if the birth rate has proven to deviate from this model (such as the birth rate being stalled) the model accounts for this by delaying the decline for between 5 and 10 years. In the case of very low birth rate countries the birth rate is increased by 0.05 children each quinquennium.

The medium model assumes the normal mortality rates. Under that assumption the current trends in life expectancy produced by the United Nations along with the effects of HIV/AIDS are used to predict the mortality rate on a country by country basis.

The final factor in the medium model is the migration level; however this can have no net effect on world population.

Figure 1 presents the United Nations statistics between 1950 and 2010, and the extrapolation to 2050. The population (diamond) is shown on the left axis and the birth rate (square) and death rate (triangle) are shown on the right axis. Under the medium model the birth rate decreases over time, the death rate dips as life expectancy is extended and then rises as that levels out. Finally, the world population grows asymptotically to about 9.2 Billion in 2050. This same 9 billion is also predicted by the US Census Bureau and the Population Reference Bureau.

9 Billion is close to the estimated maximum world population. Lutz, Sanderson & Scherbov (Lutz, Sanderson et al. 2001) estimate that there is an 85% chance that the world population will have peaked by 2100 and a 60% probability that it will not exceed 10 billion. In the next section the number of Internet users is estimated, it cannot exceed the world population at any time and therefore is unlikely to ever exceed 10 billion.

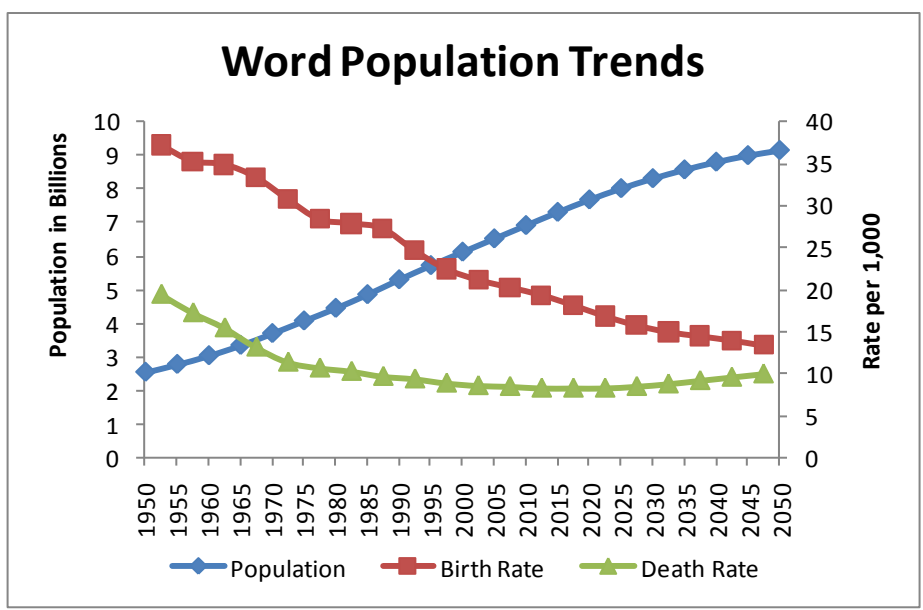

**Figure 1: World population trends between 1950 and 2050. This graph is based on the United Nations medium model which expects much of the world to reach a steady birth rate of 1.85 children per woman**

## 3.2. Growth of Internet Users

A good estimate of the number of Internet users comes from The World Bank[10]. They source the statistics from the World Telecommunication / ICT Development Report and database of the International Telecommunication Union (ITU)(International Telecommunication Union 2011). That data comes from national household surveys and Internet subscription data – and so is an estimate. The ITU defines an Internet user as anyone who has used the Internet in the previous 12 months.

Figure 2 shows the estimated number of Internet users between the years 1990 and 2009 (bottom line) along with the world population at that time (top line). At time of writing, about 26% of the world population has accessed the Internet in the previous 12 months.

The current uptake rate is about 3% per year. The ITU considers this to be too slow and identifies several problems with uptake including: basic economic issues (such as the cost of a computer); monopolies on the backbone; the lack of content in the local languages; and the level of education necessary to use it.

Predicting the future uptake is particularly difficult because of the current spread of Internet enabled mobile phones (smart phones). The ITU are predicting a sharp increase in uptake but do not go so far as to make any claims as to what that might mean.

For the purpose of the investigation herein the number of Web users is assumed to grow following a sigmoid (actually, cubic) function from nearly 0 in 1990 to the world population in 2050 (shown as a solid line in Figure 2). This choice was made because such a model fits the historic data while also preventing the possibility of greater than 100% uptake. Indeed there could well be a spike in uptake due to the mobile Internet; however there is no historic data that could be used to model such a spike. It is also noted that the diffusion of innovation model (Rogers 1962) suggests that new technology spreads through a population in a normally distributed way. The cumulative sum of a normal distribution is sigmoid.

---

[10] www.worldbank.org

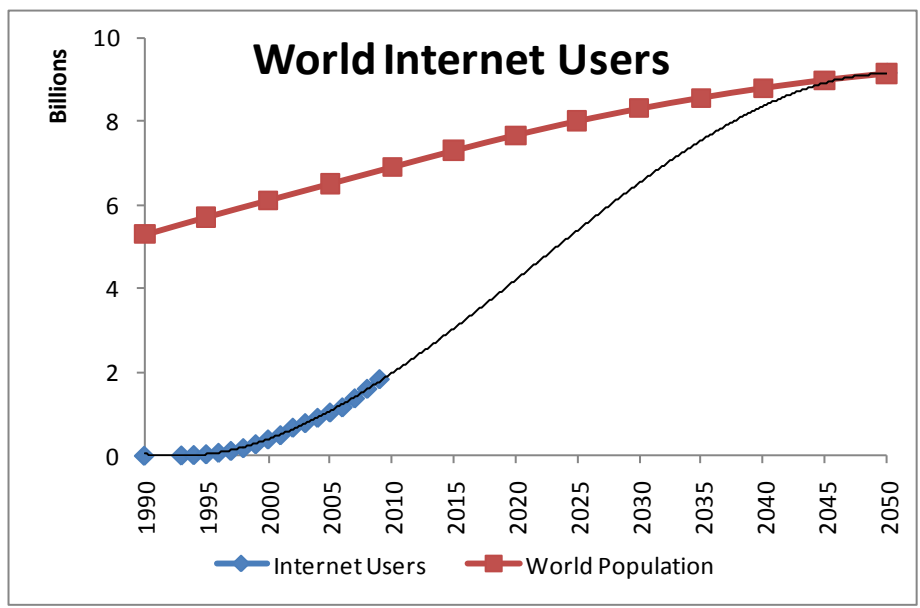

**Figure 2: Number of Internet users between 1990 and 2010. Based on numbers published by the World Bank**

The assumption of full uptake at 2050 is arbitrary. 2050 was chosen as that is about the time that the population stabilizes due to the uptake of other social infrastructure. It is reasonable to assume that full Internet uptake will not occur until at least the date at which other basic services (such as food and contraception) are up-taken.

It is also unlikely that there will ever be 100% uptake. The infant population, for example, who cannot yet speak will find the Internet of little direct use. The infirm, the illiterate, those who deliberately choose a technology free lifestyle (and so on) are also unlikely to be users. Illiteracy levels are high, UNESCO predicts the world literacy rate for those over the age of 15 will have risen to only 85% by 2015.

Data similar to that of the World Bank data is also published by Internet World Stats (Miniwatts Marketing Group 2010) who source it from Nielsen Online, the International Telecommunications Union, local regulators and other sources. The CIA World Fact Book (Central Intelligence Agency (US) 2011) suggests 2.1 Billion Internet users in 2010, also approximately the same number as the ITU.

This section developed a model for the number of Internet users based on historic evidence and the diffusion of innovation model. The next section predicts their searching behaviour.

## 3.3. Queries

It is important to measure the number of queries per month for two reasons. First, it gives an indication of Internet use (cf. Internet users). Second, it provides an indication of the amount of computing resource needed to support the users. It is the user behavior measured in number of queries that determines the size of the data centre, not the number of users. comScore[11] and Nielsen[12] are the two leading online measurement companies. Both provide an indication of the numbers of searches per month, the time people spend at given websites, and the search engine market share.

---
[11] www.comscore.com

The comScore measures are based on a global cross-section of more than 2 million consumers who gave them permission to confidentially capture their browsing and transaction behaviour. comScore also conducts surveys and adds consumer attitudes and intentions to their numbers, producing reports designed to help clients design marketing strategies. The Nielsen // NetRatings[13] search reporting service measures the search behaviour of users worldwide by installing real-time meters on their computers and monitoring the sites they visit. The data provided by each company is original and authoritative, but different. Herein only the number of searches and searches per user are used. For web search the most pessimistic (for the experiments herein) is the data from comScore[14] because it is consistently higher than that of Nielsen, but both show linear growth. US only data is used because it dates back to 2005 while the worldwide data only dates back to August 2007.

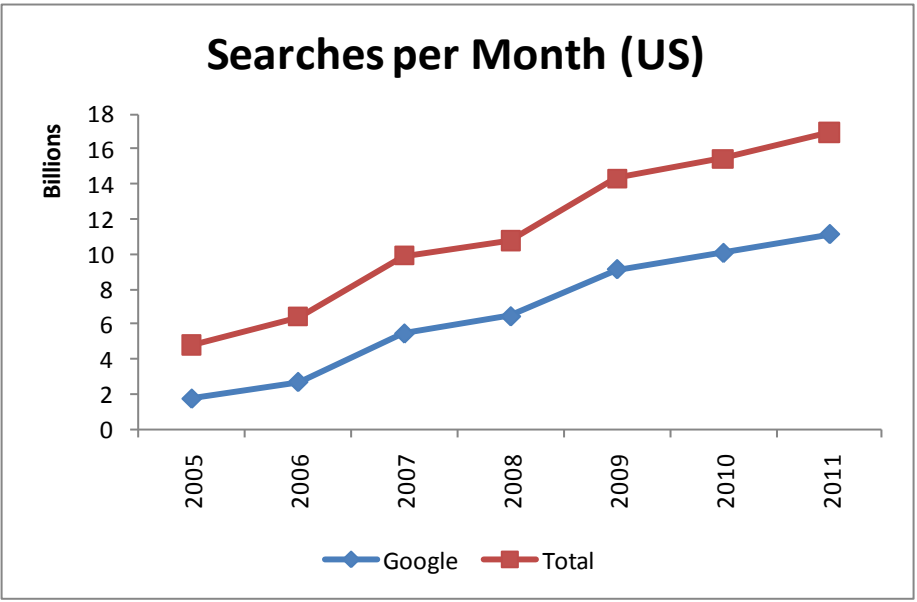

**Figure 3: Total searches per month from the US to any service provider (top) and to Google alone (bottom)**

Figure 3 presents the number of searches conducted monthly in the US as reported in the comScore Core Search Reports between 2005 and 2011 (July of 2005 and 2007, but March of all other years). This includes home and work activities for people age 15 and over as conducted at: home; work; and university locations. The numbers are based on searches conducted at the five search engines: Google; Yahoo!; Microsoft; AOL LLC; and Ask. The numbers include: local; image; news; shopping; and other vertical searches. They include all searches conducted at the search engine's domain (e.g. google.com) regardless of what is being searched for. Searches that are not on the domain of the search engine are not included (e.g. YouTube searches on YouTube.com, despite the Google ownership of YouTube). ComScore occasionally further break out the total search count into main and other sites owned by the search engine companies - in April 2010 Google saw 10.5 billion searches and all other Google owned domains including YouTube collectively accounted for 3.4 billion.

---
[12] www.nielsen-online.com
[13] Reports available from: www.nielsen-online.com/press.jsp?section=ne_press_releases#
[14] Reports available from: www.comscore.com/Press_Events/Press_Releases

Figure 3 shows that over the time period (6 years) the total number of searches has increased nearly linearly from 4.8 billion searches per month coming from the US in July 2005 to 16.9 billion in March 2011. Coincidently the number of searches per user per month is also increasing (from 80.9 in August 2007 to 103.3 in July 2009). Although it is unreasonable to project a 6 year history into a 40 year future, Figure 4 shows this projection. The lower line is Google while the upper line is the total number of searches. If the trend continues then the total number of searches expected in 2050 is about 100 billion, performed by 403 million (US) users, which amounts to about 250 searches per user per month.

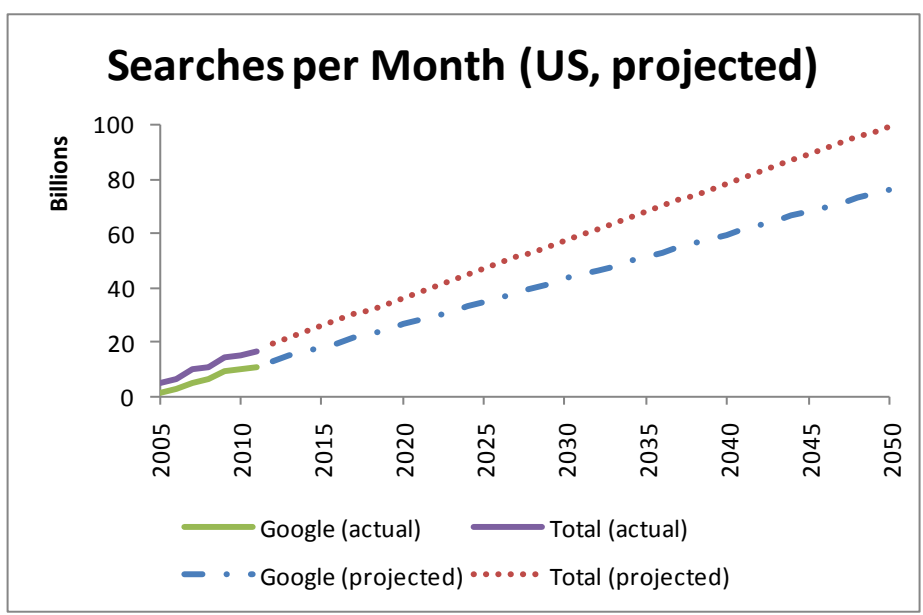

**Figure 4: Total monthly searches of five major search engines between 2005 and 2011**

### 3.4. Growth of Web Page Size

In order to predict the quantity of text on the Internet it is necessary to look at two contributing factors: the number of pages and the size of each page. This section discusses how the average web page size has changed over time and predicts the size of an average web page in 2050. Herein a page is considered to be a unique URL. Each unique URL identifies one HTML file which in turn may include several other resources including images, sounds, cascading style sheets (CSS), and script files. Human and machine generated pages are included (i.e. static pages and pages from sources such as the Wikipedia and amazon.com), but web-crawler traps and mirrors are not.

The size of the average page has increased dramatically since the inception of the web. This is for many reasons, the most obvious of which is the widespread inclusion of stylistic elements such as images. But, since web search engines only index the contents of the HTML file, it is the size of that file that is of interest.

The historical average HTML page size (excluding resources) can be calculated directly from several of the publicly available snapshots taken for research purposes.

Table 1: TREC Web Subsets

| Collection | Date | Size | Files | Mean |
|---|---|---|---|---|
| WT100g[15] | 1997 | 100GB | 18,571,671 | 5.7KB |
| .GOV[16] | 2002 | 18GB | 1,247,753 | 15.2KB |
| .GOV2[16] | 2004 | 426GB | 25,205,179 | 17.7KB |
| clueWeb09[17] | 2009 | 25TB | 1,040,809,705 | 23.5KB |
| Google 2010[18] | 2010 | unknown | 4.2 Billion | 22.3KB |

The average file size of an HTML document in the TREC collections is shown in Table 1. The first column lists the collection, the second the date of the crawl. In the third column the total size of the collection is listed, and in column four the number of documents. The final column gives the mean length of a document. For example, the clueWeb09 collection was crawled in 2009, is 25TB is size, and contains 1,040,809,705 documents; which average 23.5KB per document.

Care must be taken in interpreting these statistics. It is known, for example, that in the .GOV collection documents were truncated at 100KB. It is also known that the effect of doing so reduces the size of the collection size from 35GB to 18GB (by about half). This, in turn, halves the mean document length – however these documents are outliers. In the .GOV2 collection the documents were truncated at 256KB.

In the clueWeb09 collection care was taken to crawl documents in multiple languages (the prior collections were predominantly English). Half of clueWeb09 is English and the remainder is in Chinese, Spanish, Japanese, French, German, Portuguese, Arabic, Italian, and Korean. That is, none of the collections are truly representative of the web as it stands today let alone how it might stand in 2050.

Google release statistics on pages crawled using the Googlebot. At time of writing the most up-to-date data they provide is dated 26 May 2010, however it is not clear when the data was crawled. Google crawled 4.2 billion pages which averaged 43.91 resources per page totalling 376.67KB per page (including images, scripts, and so on). Of the 43.91 resources per pages, 29.39 were images, 7.09 were scripts, and 3.22 were style sheets. Factoring these out of the totals leaves 4.21 resources totalling 93.98KB per page. Assuming each of these is of equal average size (because it is not clear what they are) results in an average of 22.32KB per resource; approximately the same value as the mean clueWeb09 HTML document size.

Care must be taken interpreting the Google results. Not only are the 4.21 resources of an unknown type, but Google crawls the web in priority order, and follows robots.txt file instructions. Most importantly, the web crawler identifies itself as being a crawler and so may be presented with a different document than that which a user sees when visiting the same page.

The growth of HTML page size (from Table 1) is shown in Figure 5. The growth has been nearly linear since 1997 with the mean growing at the rate of about 1.5KB per year.

If the current trend continues then by 2050 the average size of an HTML document will be about 76KB. That is, three times today's size. Given a single page of A4 holds about 500 words, which at 5 characters per word is between 2.5KB and 5KB per page (depending on character encoding in 8 or 16 bits), 76KB appears to be excessive (30 pages). Indeed, the 30 page document of 2050 will be a fundamentally different object to interact with than the 2 page document of 1997. This warrants further investigation, however it seems reasonable to use the number as an overestimation.

---
[15] See Hawking & Craswell (1998)
[16] http://ir.dcs.gla.ac.uk/test_collections/
[17] http://boston.lti.cs.cmu.edu/Data/clueweb09/
[18] http://code.google.com/speed/articles/web-metrics.html

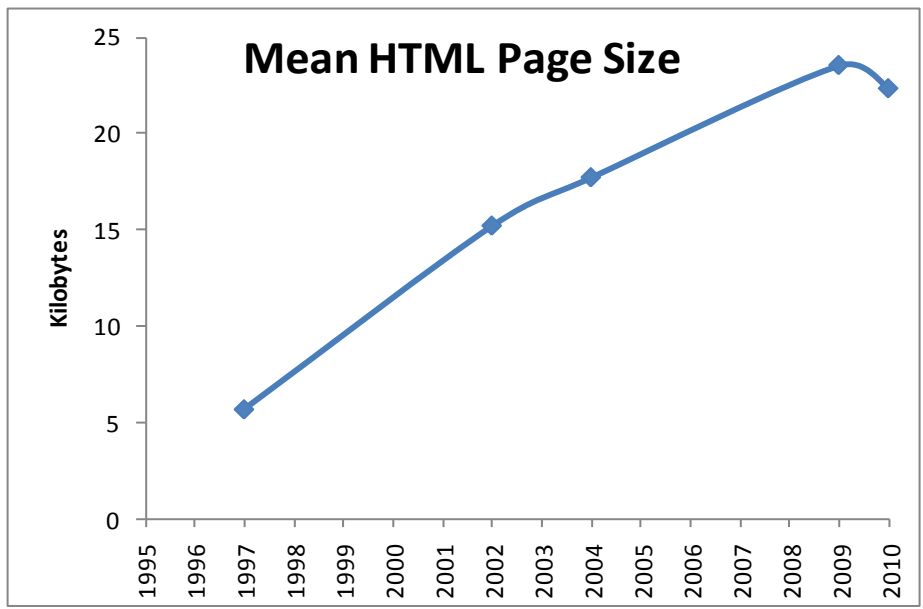

**Figure 5: Mean size of an HTML page based on the TREC crawls between 1997 and 2009, and the stats published by Google in 2010. The average page size is increasing at a rate of about 1.5KB per year**

An alternative way to predict the increase in page size is to extrapolate from Wikipedia page size history. This information is provided online by the Wikimedia Foundation (Wikimedia Foundation 2011). Figure 6 shows that the web trend of increasing page size is also seen in the Wikipedia. Between 2001 and 2009 the average size of an English (en) page increased from 1064 bytes to 3455 bytes, by a factor of 3.2. For the whole Wikipedia it increased from 1048 bytes in 2001 to 2534 bytes in 2009, by a factor of 2.4. Extrapolating to 2050, the average English page will be nearly 15KB and the average page across all languages will be 9.6KB. That is, an increase of a factor of 4.4 for English and of 3.8 overall (from 2009). Wikipedia pages are substantially smaller than web pages because the wiki mark-up language is more compact than HTML.

The Wikipedia numbers are in line with those for the web. Pages are getting longer at a linear rate and if they continue to do so then they can be expected to be about three times today's size in 2050.

### 3.5. Growth of Number of Web Pages

This section examines the trend of the growth rate of the number of web pages. Estimating the size of the web is fraught with problems for many reasons. The size of the hidden web (for example, password protected pages) cannot easily be estimated, and with so-called Web 2.0 it is hard to define what, exactly, a page is. There are also problems with replication, duplication, spam, and so on. This section starts with estimates with respect to the Wikipedia where such problems are minimal, and then examines the web as a whole.

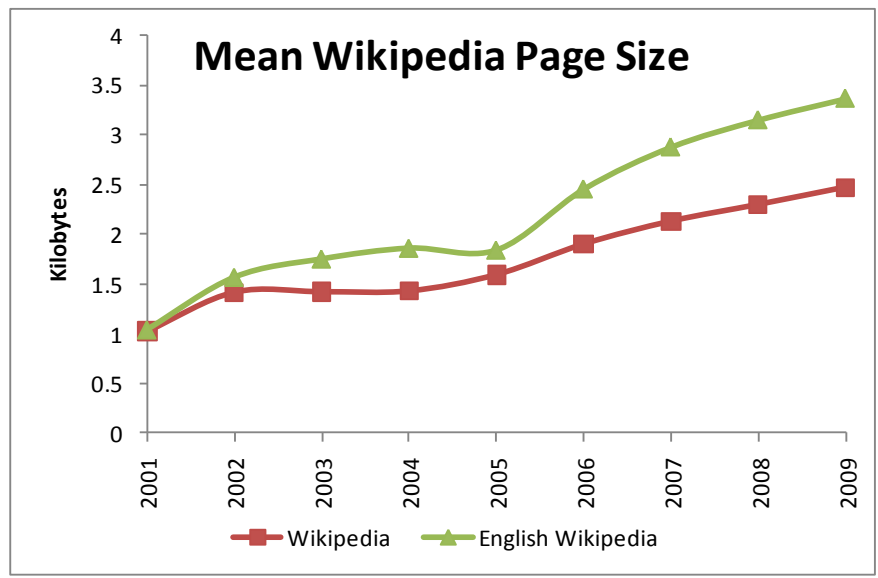

**Figure 6: Wikipedia page size in kilobytes between 2001 and 2009 for the whole Wikipedia and for the English (en) Wikipedia shows the same trend as for all web pages**

### 3.5.1 Wikipedia

A Wikipedia page is clearly definable. Each page is on a unique topic, no topics are duplicated, and none are spam. It serves as an excellent model for those parts of the web that a search engine should index. The growth statistics are also publically available.

Figure 7 shows, on the left axis, the number of articles in the Wikipedia in both English (bottom line) and all languages (middle line) drawn from statistics published by the WikiMedia Foundation (Wikimedia Foundation 2011). In the early years of the Wikipedia the number of documents was growing at a phenomenal rate; however, since 2006 the growth rate has been stable and linear with about 2.9 million documents in total being added every year.

The pattern for the English Wikipedia is more interesting and is shown in the figure (top line, right axis). After an initial very rapid growth rate, the rate stabilized at about 600,000 new articles per year. Since 2007 the growth rate has been falling by about 100,000 articles per year.

The organization of an encyclopaedia such as the Wikipedia allows for only one article per topic. To maintain exponential growth there must be an inexhaustible number of topics to include. In practice there is a finite number of topics. The most important (to the audience) topics are added first with the less important topics being added afterwards. Sooner or later the remaining pool of topics becomes so obscure that few editors of the encyclopaedia would be expected to be able to add an article on that topic. This is why the English Wikipedia is showing a decline in the growth rate, it is nearing saturation.

Saturation is happening first in the English Wikipedia because it was the first language of the Wikipedia and Internet penetration is high in predominantly English speaking countries. It has not occurred in all languages – and hence the overall growth rate is not in decline. High growth rates are seen in languages such as Vietnamese, Burmese, and Indonesian.

It is reasonable to expect continued high linear growth in the Wikipedia for many years because it will take many years before all language versions saturate (at time of writing, for example, the Marshallese language version has only 11 articles whereas English has 3,628,406). Eventually all language versions will saturate. At that time new articles will be added for new events and discoveries, old articles will continue to be edited and extended, but few articles will be deleted.

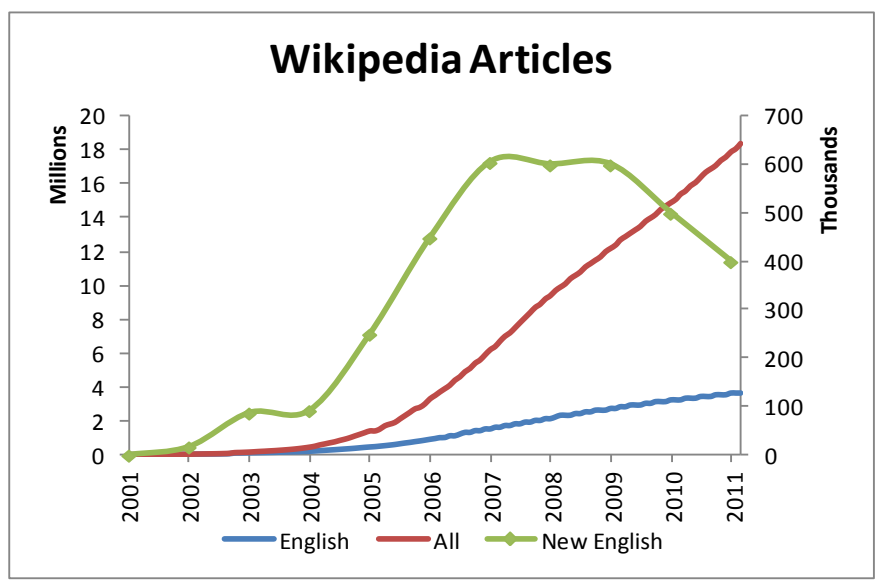

**Figure 7: Number of Articles in the Wikipedia in all languages (top) and English (middle). Also included is the number of pages added per year (top line, right axis)**

3.5.2 Web

A search engine can only crawl surface (visible) web pages, that is, those pages that are not hidden or password protected. Information on the deep (or hidden) web, including those that are not linked and can only be found in response to a local site search are generally inaccessible to Web crawlers. Google's deep web surfacing system (Madhavan, Ko et al. 2008) does find some deep web resources; however it finds only a very small part of the deep Web. The deep Web remains largely unexplored by the search engines (He, Patel et al. 2007).

In 2005 Gulli & Signorini (Gulli and Signorini 2005) conducted a study that used the search engines to estimate the size of the web. They conducted searches in 75 different languages and estimated that there were over 11.5 billion web pages in the publicly indexable web at that time.

Neither Google nor Bing report the number of pages in their archive. Cuil, however did. Cuil was launched in July 2008 with 121,617,892,992 web pages. According to Muchmore's popular-press article[19], Cuil founder Anna Patterson claimed that the archive was the largest of any search engine and nearly three times the size of Google's. Cuil was reportedly only 20 billion away from the then estimated 141 billion pages on the visible web. In response, Alpert & Hajaj announced on the Google blog that Google had discovered one trillion unique URLs (Alpert and Hajaj 2008).

Cuil's meteoric rise was followed by a sudden crash. This may be because it favoured recall (whereas Google favoured precision), the same mistake made by Alta Vista. In September 2010, shortly before going offline, Cuil reported that the archive contained 127 billion pages, much of which was presumably spam.

Snapshots of the whole web at various moments in time are necessary for estimating the growth rate of the web. Such snapshots exist indirectly in the Internet Archive's Wayback Machine (Internet Archive 2011). The Internet Archive is a non-profit organization which has built a digital library of Internet sites and other digital cultural artefacts. The Wayback machine is an archive of old web pages. It is fascinating to use the site to see how various well known web sites have changed their look and feel since they started.

---
[19] www.pcmag.com/article2/0,2817,2326741,00.asp

Figure 8 shows the number of pages included in the Wayback Machine between 2002 and 2010. In 2002 there were 30 Billion pages, it grew in steps and in 2008 reached 150 billion. The number reported at present continues to be 150 billion.

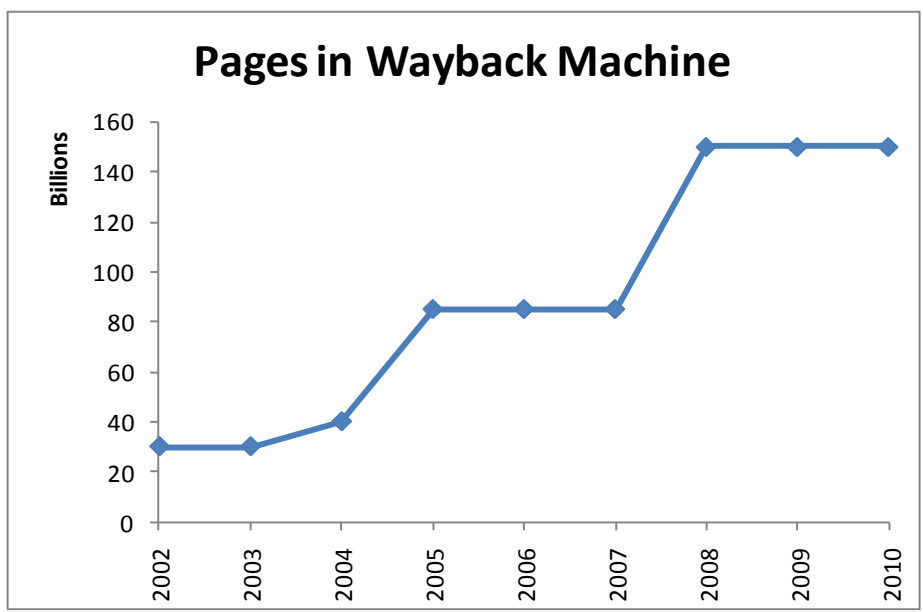

**Figure 8: Documents in the Internet Archive's Wayback Machine. This includes current pages and old versions of those pages as well as defunct web sites**

From the statistics in the Wayback Machine and the Cuil vs. Google debate it appears as though in 2008 there were about 150 Billion pages in the visible web. A substantial portion of these pages are not generally considered helpful for resolving any information need. Such pages include spam, phishing pages, and duplicates of various resources (such as the Wikipedia and Yahoo!).

Commercial search engine companies go to tremendous lengths to filter out spam and otherwise non-useful pages. They index only what they consider to be the important or interesting pages. As they focus on precision, a large index does not necessarily mean better results (the mistake Cuil made). However, it is necessary to index enough pages to be able to answer all possible information needs, no matter how obscure.

Historic information on the size of the Google archive between 2001 and 2005 is available on the Wayback Machine and is presented as the left hand line of Figure 9. Google started divulging numbers on or about 11 July 2000 and reported 1,060,000,000 pages, then 8,168,684,336 at the end of 2004. This number remained constant until early 2005 and then Google stopped reporting numbers. On 26 September 2005, its 7[th] birthday, Google announced on their blog that they are offering users "a newly expanded web search index that is 1,000 times the size of our original index"[20]. This suggests that by 2005 Google had indexed about 26 billion documents as the first

---

[20] http://googleblog.blogspot.com/2005/09/we-wanted-something-special-for-our.html

1998 Google index had 26 million pages[21]; the Google prototype had 24 million pages (Brin and Page 1998).

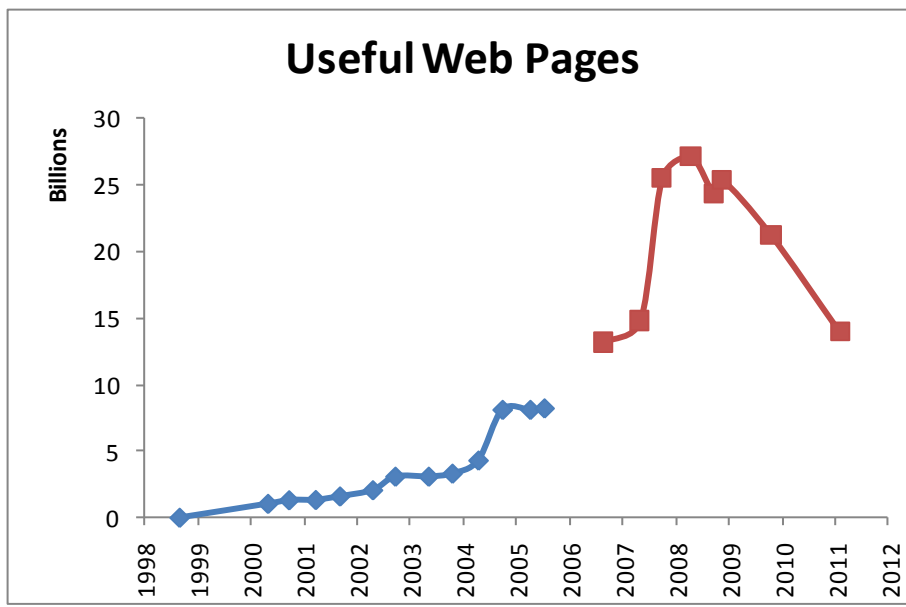

**Figure 9: Pages indexed by Google according to Google (left in blue) and the estimates of the number of indexed pages on the web (right in red) according to de Kunder. The discontinuity is as a consequence of deKunder starting to report only after Google stopped reporting.**

De Kunder (de Kunder 2006) estimates the size of the (useful) web by sampling the indexes of Yahoo!, Google, Bing and Ask. The relative frequency of words in a natural language follows Zipf's law. By taking a large enough snapshot of natural language these frequencies can be measured. De Kunder chose a one million page dump from DMOZ[22] and 50 words that were evenly spaced across the Zipfian distribution. Each day these 50 words are submitted to the search engines and the number of documents reportedly containing these words is recorded. By simply dividing the reported count by the proportion of documents expected to contain the word (from Zipf's Law) the true count of pages containing that term is given. These counts are averaged over the 50 words giving an estimate of the true size of the archive indexed by each search engine. De Kunder's result includes not only HTML pages, but any content the search engine purports to index including: PDFs, news stories, videos, products, spreadsheets, and software in the Apple and Android app-stores.

To estimate the size of the web de Kunder takes the top results from each search engine and computes the overlap – the reciprocal of this is the uniqueness of each search engine. These are then combined to a final web size count.

As de Kunder identifies, the final result is dependent on the order the overlap is computed. Herein the YGBA numbers are reported. These are the results of starting with Yahoo!, then subtracting Google, Bing, and finally Ask. The YGBA results are used because they are reported in the text of de Kunder's web site[23] which are stored in the Wayback Machine and from which it is possible to retrieval the scores.

---

[21] http://googleblog.blogspot.com/2008/07/we-knew-web-was-big.html
[22] http://www.dmoz.org/

The results are shown on the right of Figure 9. The estimates start shortly after Google stopped reporting numbers and believably extrapolate the Google numbers. The dip in recent years is not the web getting smaller, but is more likely to be the spam filters getting better or the search engines converging on the same set of top documents.

Care must be taken in interpreting the numbers as they are only estimates. Google, for example, released a word list showing that their English index contains over 13 million terms (Lam 2010), de Kunder used only 50.

De Kunder's estimate is based on a sound theory, and it is also consistent with reports in the popular press. For example, in 2005 Yahoo! claimed to provide access to over 20 billion items which include "19.2 billion web documents, 1.6 billion images, and over 50 million audio and video files" (Mayer 2005). This is over twice as many documents as Google which, at the time, contained 8,168,684,336 pages[24]. Over the last two years Google appears to have peaked somewhere between 40 and 45 billion pages whereas Yahoo! peaked somewhere between 60 and 65 billion pages. The drop to just over 10 billion can be seen when Yahoo! adopted Bing in 2009 (however Bing appears to remain smaller than Yahoo!) (Costa 2011).

Figure 9 as a whole shows that Google's 1998 25 million web page prototype had grown to over a billion pages in 2000. It then roughly doubled every two years until 2004 and in 2005 jumped to about 8 billion but maintained the doubling every two years. Since 2006, estimates as to its size suggest the archive has grown slowly (at best). The introduction of the Google Caffeine indexing system did not see a huge jump in page numbers, an observation confirmed by Matt Cutts (head of Google's Spam team)[25]. At present de Kunder's estimate for the size of Google is between 34 and 35 billion pages, however this is not only an estimate, and not only larger than the estimate of the whole web (14 billion), but is estimated with the knowledge that Google itself estimates the number of documents containing a given term – and those estimates fluctuate wildly[26].

The size of the indexed web can believably decline over time for many reasons. As the search engine companies improve their spam and duplicate detection technology an increasing number of pages will be rejected from the index. When web sites become defunct their pages will be removed.

### 3.5.3 The Estimate

To estimate the number of pages that will exist in the future it is necessary to model how pages were created in the past. As an estimate of the total number of pages on the web has already been given in section 3.5.2, the missing component is the number of human Internet years that it took to generate those pages.

Herein a human Internet year is defined as one Internet user in one year. That is, if a given individual is a user of the Internet for two years and a second user is a user for one year then together they represent three human Internet years. The statistics for the number of Internet users are given in Section 3.2. A running total at the end of each year gives the total number of human Internet years.

---

[23] http:/www.wordwidewebsize.com/
[24] Google webpages archived in the Wayback Machine (itself at: http://www.archive.org)
[25] www.seroundtable.com/archives/022372.html
[26] http://ilk.uvt.nl/events/dekunder.html

Figure 10 shows the number of pages that existed each year divided by the running total of human Internet years. The line on the left (2000-2005) is based on the number of documents Google claim to have indexed divided by the World Bank estimates of the number of Internet users, while on the right (2006-2009) the numerator is the WordWideWebSize (de Kunder's) estimate and the denominator the World Bank estimate. There is a discontinuity because de Kunder started estimates after Google stopped published statistics.

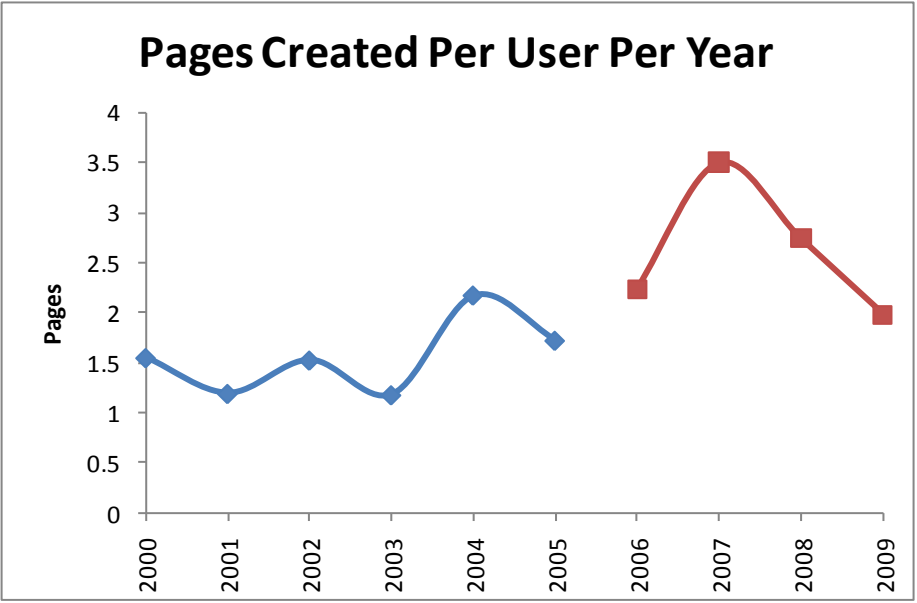

**Figure 10: Total number of pages created per human Internet year so far on the Internet. On the left (in blue) is the numbers based on Google's quoted archive size while on the right (in red) is the number based on the WorldWideWebSize estimated. Since WorldWideWebSize started reporting only after Google stopped there is discontinuity**

The figure shows that the creation rate was stable between 2000 and 2005 at about 1.6 documents per person per year. Between 2006 and 2009 it was not stable, but the most recent estimate is back in line with the earlier estimate, being about 1.9 documents per person per year. It is reasonable to expect the document creation rate to be at most constant per person per year. There is a maximum rate at which any individual can create a new document on a new topic – they are limited by their typing rate.

What is unexpected is that it *is* (roughly) constant. Section 3.4 shows that pages are becoming longer over time. On the assumption that users spend constant time editing pages, it is not possible for them to spend that time both extending pages and creating new long pages. It is possible that late web adopters spend more time editing than the early adopters, but this hypothesis has not been tested herein. An alternative explanation is that the pages might contain non-unique or non-human generated content (such as JavaScript). However, if this were the case then Wikipedia articles would not be increasing in length – which they are (see section 3.4).

Assuming the Internet user growth rate from Figure 2, and a constant creation rate of 2 pages per user per year (in excess of that in Figure 10), Figure 11 shows how many pages can be expected to be in a search engine archive between 1990 and 2050. The values used prior to 2010 are observations while those after 2010 are projections. From the figure, about half a trillion web pages will be indexed by the search engine companies by 2050.

This section examined the number of web pages expected to be in a search engine archive. It first examined the Wikipedia and showed that its English version has reached topical saturation. It then examined the web and using factual and estimates of search engine archive sizes between 2000 and 2009, and user counts over the same period, showed that there is a nearly constant creation rate of indexable web pages – that rate being just below 2 pages per person per year. From this a projection of the number of indexable pages expected to exist between 2011 and 2050 was made.

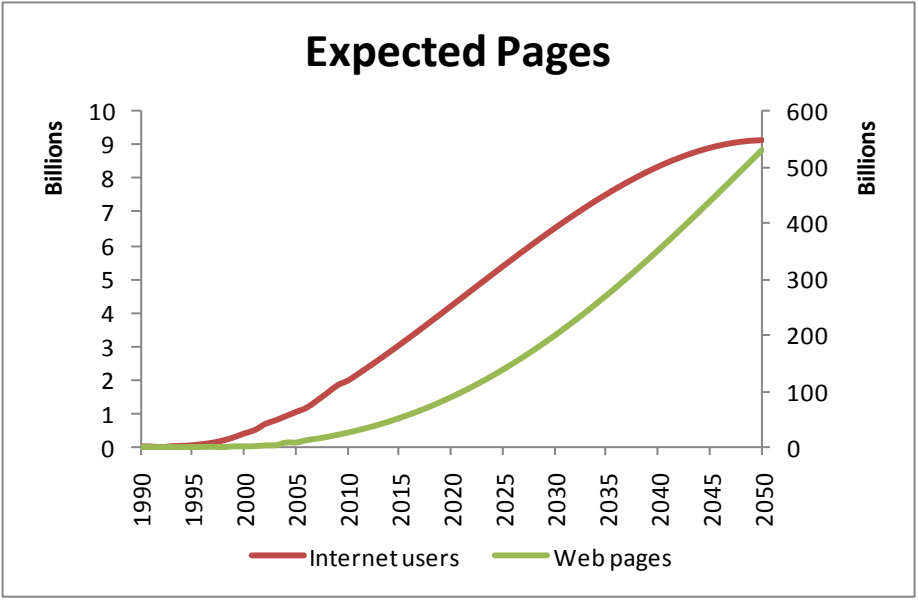

**Figure 11: Estimated number of Internet users (top line, left axis) along with the number of web pages that are expected to exist (bottom line, right axis). Numbers between 1990 and 2010 are observed values while those between 2011 and 2050 are projections**

### 3.6. Number of Web Sites

The previous section examined the growth in the number of web pages over time. It accounted for both those pages that are created and those that are deleted. No account was made for the number of web sites. Although this is also hard to estimate, for the purpose of illustration (as these statistics are not used beyond this point) web site numbers are equated with domain names herein.

Each quarter VeriSign publish a brief containing various statistics including the total number of registered domains[27]. For the combined .net and .com domains the VeriSign reports include the renewal rate. The rate has remained essentially stable since the reports were first published (2004) at between 70% and 77%. The other nearly 30% of the domains are left to expire. Figure 12 shows the total number of registered domains worldwide alongside the renewal rate for .com/.net domains between 2004 and the end of 2010. There is a steady nearly linear increase in the number of domains over time while the renewal rate remains nearly constant. It is reasonable to believe that all top level domains (other than .com and .net) will enter such a stable period after an initial early adoption period. That is, it is reasonable to expect that about 30% of the domains due to expire do expire.

---

[27] See both current and archived "Domain Name Industry Brief" at:
www.verisigninc.com/en_US/why-verisign/research-trends/domain-name-industry-brief/index.xhtml

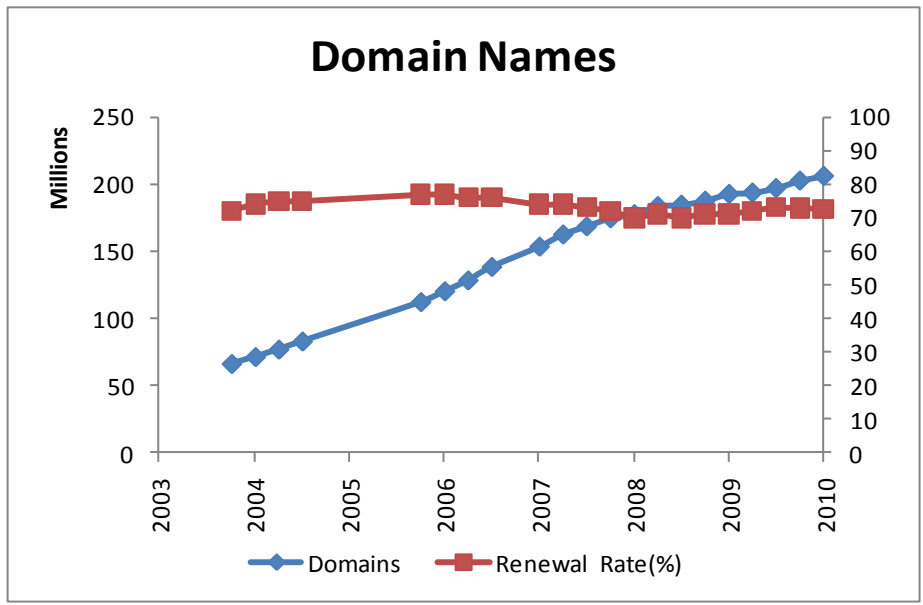

**Figure 12: Total number of unique domains worldwide and the renewal rate of .com/.net domains. The number of domains continues to grow, but the renewal rate is essentially constant in the low 70%**

O'Neill *et al*. (O'Neill, Lavoie et al. 2003) examined the growth rate and attrition rate of the public-web. They define a public-web site as a site that offers the majority of its content to the public on an unrestricted basis. Their method involved randomly probing a large proportion of the IPv4 address space and crawling content therein. Using this method they observed an attrition rate of 17% between 2001 and 2002.

The VeriSign reports and the work of O'Neill *et al*. cover different time periods and so it is not possible to reconcile the difference. They also present subtly different values. The VeriSign score is all domains whereas the number of O'Neill *et al*. is for IP addresses that resolve to public-web sites. VeriSign do observe that domains that resolve to web sites are more likely to be renewed than to expire, and present statistics on the proportion of domains that are only 1 page (parked), more than 1 page (web sites), or are not web sites. Of the 104.6 million domains sampled in January 2011, 21% were parked, 67% were web sites, and 12% did not contain web sites.

If the current linear growth rate continues then by the end of 2050 there will be over 1.2 billion web sites, just over 70% will be renewed each year.

### 3.7. Growth of Index File Size in Search Engines

In this section an estimate of the size of the inverted index necessary to search the indexable web is made. This estimate is based on the projected number of web pages, the average size of a web page (from previous sections), and an estimate of the size of the index relative to the size of the document collection. Figure 11 has already presented the history of the number of web pages and a model for future growth based on population size and creation rate. Figure 5 has already presented the history of the size of a web page. The growth was linear over time and predicted to remain linear into the future.

An estimate of the total indexable web size is simply the number of documents multiplied by the average document size. This is shown in Figure 13 for both the historic and future data. The projection suggests that by 2050 there will be about 37 Petabytes of text in half a trillion pages averaging 76 Kilobytes each (and served from 1.2 billion sites).

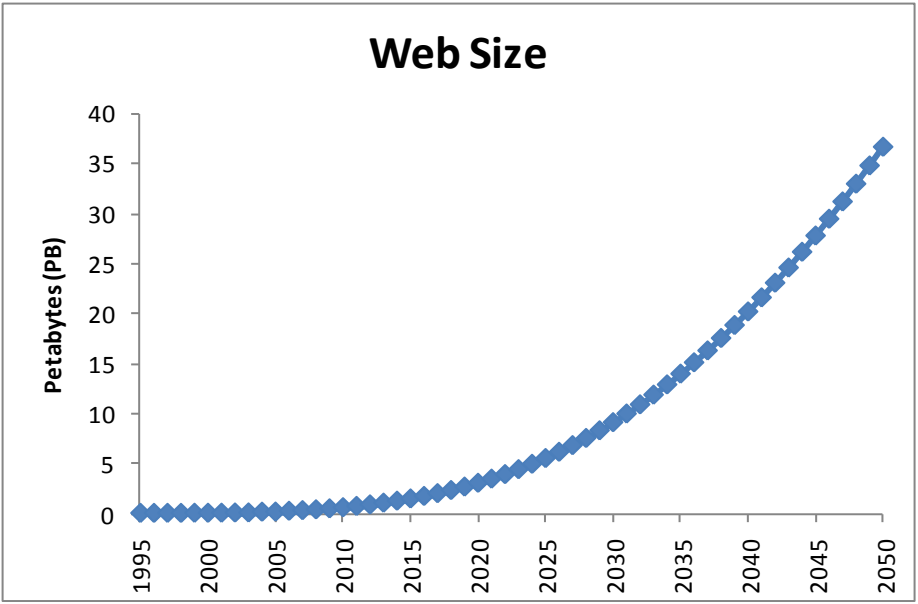

**Figure 13: Estimated historic and projected future size of all web pages. By 2050 the total size is estimated to be about 37 Petabytes (PB). For comparison, the Wayback Machine currently contains about 150 billion pages totalling 2PB**

Grime (Grime 2010) states that the Google Caffeine system stores nearly 100 million gigabytes (100 Petabytes) of data in one massively distributed database. This database is already over twice the size necessary to store all the web pages expected by 2050. But, as discussed in Section 3.4, the web page size is a small fraction of the total content displayed to the user; that includes, amongst other content, multimedia objects – which Google also stores and indexes.

An estimate of the size of the search engine inverted index necessary to index the 2050 indexable web can be estimated by sampling index sizes of large collections of web pages.

The size of an inverted index is influenced by many factors ranging from compression scheme to whether or not phrase and proximity search is included (and how). For the experiments herein a term only (without word positions) inverted file search engine that stores the postings TF-impacted and compressed with difference encoding and variable byte encoding was used. The search engine, ATIRE, does not perform index pruning or stop word removal. This search engine was developed at the University of Otago and has proven to be both efficient and high precision at whole-document retrieval at INEX (Trotman, Jia et al. 2009).

Table 2 shows the size of the inverted index for two standard test collections: the ClueWeb09 Category B collection and the INEX Wikipedia collection. Column 1 lists the collection, column 2 lists the number of documents while column 3 lists the total size of the collection. Column 4 lists the size of the inverted index while the final column lists the relative size of the index to the text. For example, the ClueWeb09 Category B collection has 50,220,423 English documents totalling 1.5TB. The index of the collection is 32GB in size which is 2% of the collection size. For the INEX Wikipedia collection, the index size is 3% of the collection size.

**Table 2: Collection size and index size for two standard IR test collections suggest that the index is about 2% of collection size**

| Collection | Documents | Size | Index | Ratio |
|---|---|---|---|---|
| ClueWeb09 (B) | 50220423 | 1.5TB | 32GB | 0.02 |
| INEX Wikipedia | 2666190 | 50.7GB | 1.4GB | 0.03 |

Prior work has suggested the index can be expected to be larger than that in Table 2. For example, the Indri index for ClueWeb09 Category B is 174GB[28], or 11% of the collection size. Indri presumably supports more features than the ATIRE search engine and hence the index is larger. With today's technology the index of the 37PB web of 2050 would range between 4PB (at 11%) and 733TB (at 2%). Hereinafter the former number is used (11%) because it is pessimistic.

Section 3 developed a model for the size of the web. That model included the user base, the growth in the number of the documents and the growth of documents themselves. Finally an estimate of the index was made. It is estimated that by 2050 there will be half a trillion indexable pages and the index will be 4PB (at 11%).

## 4. Growth in Storage Capacity

In the previous section estimates of the size of the indexable web, and its index, are made. In this section the history of storage device capacity is examined and future projections are made. Finally, the section shows that the two cross – that is; it will eventually be possible to store the entire indexable web on a single storage device.

### 4.1. Growth of Hard Disk Capacity

In the previous section an estimate is made of the size of an index of all static pages on the web. By 2050 that index will be between 733 TB and 4 PB.

The capacity of hard disk drives purportedly follows Kryder's Law (Walter 2005). This law states that capacity doubles every year. By comparison, Moore's law states that the number of transistors on a silicon die doubles every 18 months. The best way to determine hard disk capacity growth rate is to measure it.

The first hard disk drive was marketed by IBM in 1956 for use with the IBM 305 RAMAC. The drive, IBM 350 Disk File, had 50 platters each 24 inches in diameter and stored a whopping 5 million characters[29]. Characters at that time were 6 bits plus a single parity bit.

Personal Computers were being shipped with hard disk drives by March 1983 when IBM released the PC XT. The XT included a 10 Megabyte ST-412 from Seagate. Just over a year later, August 1984, IBM released the IBM PC AT with a 20MB hard drive. Two close sample points is not sufficient to claim a trend.

In September 1981 Apple released the 5MB ProFile hard disk drive for the Apple /// (and later for the Apple ][ and Lisa). Internally the ProFile included a Seagate ST-506. In 2011 Amazon.com offers, for under US$200, 3TB hard disk drives from both Hitachi and Western Digital. These two sample points are 30 years apart and better for estimating the trend: 600,000 times increase in capacity over 30 years.

---

[28] www.lemurproject.org/clueweb09/indri-howto.php
[29] http://www-03.ibm.com/ibm/history/exhibits/storage/storage_350.html

Nienhuys[30] crawled the web for information purporting to state advertised hard disk capacities. His data is available from Wikipedia[31] and is presented in Figure 14. The date is on the horizontal axis and size in Terabytes is on the vertical axis with advertised capacities represented as check marks. Starting from 5 MB in 1980, the dashed line is the capacity predicted by Kryder's Law and the solid line is that predicted by Moore's Law. Commodity hard disk sizes appear to follow Moore's law, doubling in capacity every 18 months over the last 30 years. If this trend continues then commodity hard drives in excess of 300 Exabytes will be on the market by 2050.

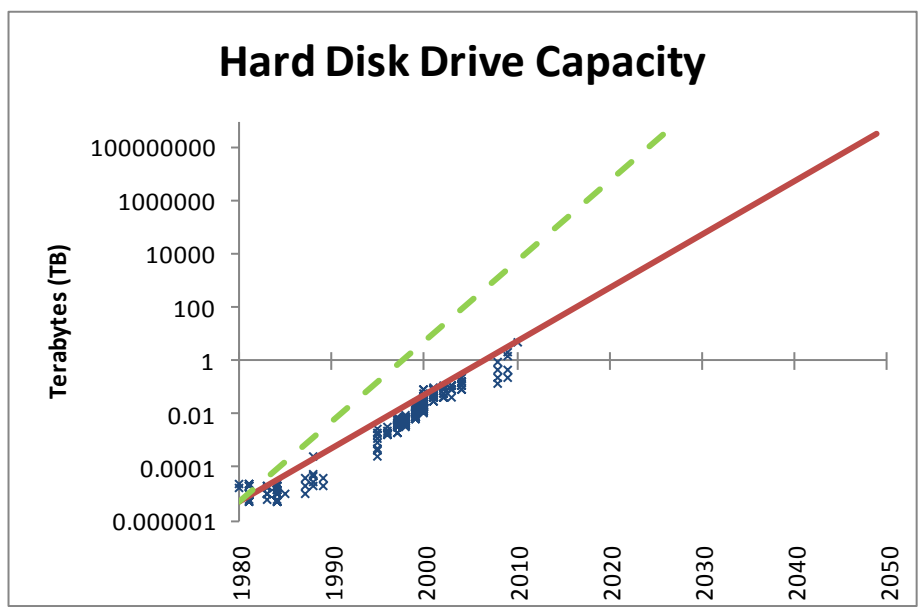

**Figure 14: Advertised commodity hard disk drive capacity (crosses). The dashed line is that predicted by Kryder's Law and the solid line is that predicted by Moore's Law**

By overlaying the projected growth rate of hard disks and the size of the inverted index of the web (at a pessimistic 11% of the collection size) it becomes obvious that at some moment in time the entire index will fit on a single commodity drive. In Figure 15 hard disk drive capacity prediction is shown with squares and index size is shown as triangles. The cross over point is shortly before 2020 – that is, within the current decade. Also shown in the figure is the projection of the size of the textual part of the indexable web. That will fit on a hard drive shortly after 2025 with room to boot (literally). Certainly before 2030 the text, the index, the software to search it, and the operating system can be expected to fit on a commodity hard disk drive shipped with an off-the-shelf desktop computer.

This discussion does not account for images on those web pages, videos, sound, and so on. According to the Google survey in 2010 (Ramachandra 2010) the average size of a web page on the wire (including all components with some compressed) was 320.24KB and the uncompressed document size was 376.67KB, over 10 times the size of the text. With hard disk drive capacity doubling every 18 months one must wait only 6 years before the capacity has increased 16 fold. That is, if the sum of the components remains at 14 times the size of the text then within 6 years of

---

[30] http://commons.wikimedia.org/wiki/File:Hard_drive_capacity_over_time.png
[31] Nienhuys' data is not from a verifiable source, and is used here only to fill-the-gaps between the 5MB and 3TB samples already presented

2025 it will be possible to place not only the index and the text, but also all the images, video, sound, and so on a single commodity hard drive.

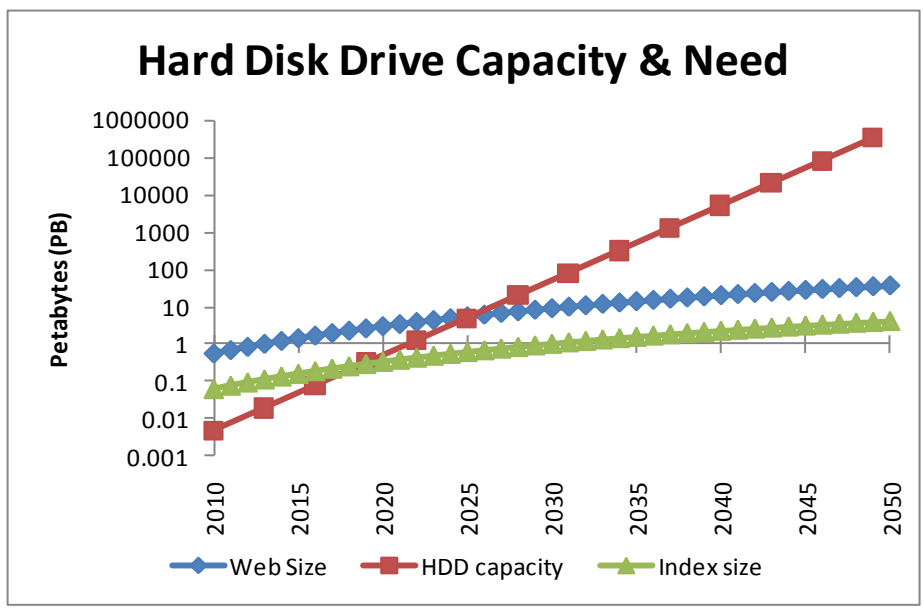

**Figure 15: Predicted hard disk drive capacity (squares) crosses the web predicted index size (triangles) before 2020 and the size of the web's text (triangles) before 2030**

It is acknowledged that these estimates are with respect to the indexable web only. Search providers typically offer multiple products. Google indexes and makes searchable, for example, such data as street maps (Google Maps).

This section examined the growth of commodity hard disks between 1980 and 2011. They increased in capacity 600,000 times in that period. If they continue to follow Moore's law then shortly after 2030 the indexable web and its index will fit on a single drive.

### 4.2. Growth of Memory Card Capacity

This section examines the growth of portable device storage capacity in an effort to predict when it will be possible to carry the entire indexable web on a smart phone.

Walter (Walter 2005) suggests that hard drives will continue to miniaturize until the point at which they eventually replace low-capacity flash memory cards. When this happens they will migrate (back) into consumer devices such as phones, cameras, and PDAs. In the mean time, the storage capacity of a mobile device is limited by the size of solid state flash drives. SD cards currently hold a 78% market share and as such are the de facto standard in flash memory (SDAssociation 2009); and ideal for studying capacity growth.

The SD Card Association (Lin 2009) plot the history of card capacity between 2000 when they were introduced until 2009. Panasonic (Lai 2010) provide historical data for capacities between 2005 and 2009. Lexar[32] are currently marketing 128GB SDXC cards and microSD cards of 32GB.

Figure 16 shows card capacity growth between when they were introduced in 32 and 64 MB capacities through to the current 128GB cards. Over this period capacity has doubled every year (ironically, following Kryder's Law not Moore's Law). The current SDXC card specification includes cards of up to 2TB in capacity, but these are not currently available (they will be by 2025 at the current growth rate).

---
[32] www.lexar.com/products/lexar-professional-133x-sdxc-card?category=4155

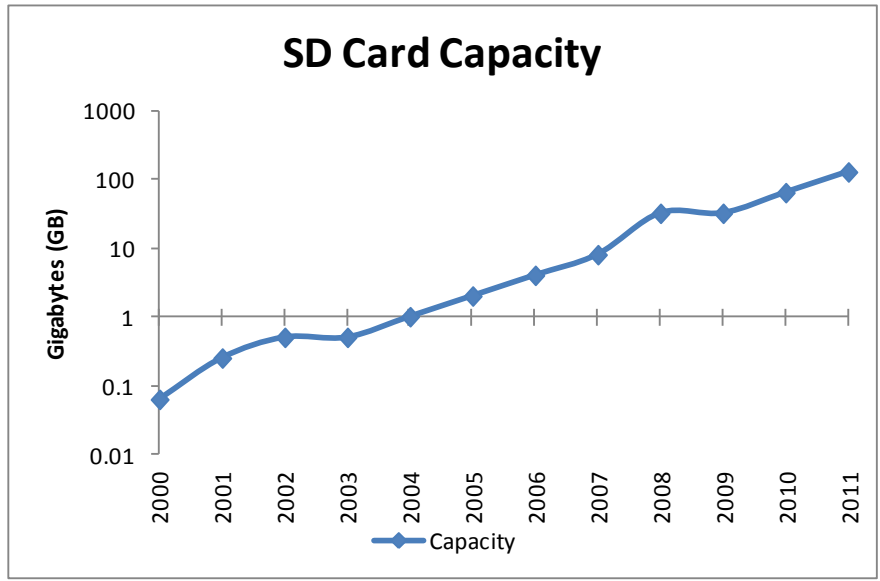

**Figure 16: SD card capacity between 2000 and 2011**

The capacity of memory card depends on miniaturization of transistors. This, in turn, relies on doping which relies on photo lithography. Consequently it will become increasingly difficult to make such devices over time (Lai 2008). The miniaturization problem can be overcome by making larger dies, 3D dies, or by placing more than one die on the same card. For the purpose of this discussion capacity is not considered to be constrained by miniaturization.

Overlaying the growth rate of the SD card with the indexable web index size suggests that the index will fit on a single card by 2035 when the index is over 1PB and the card is 2PB. By 2039 all the indexable pages will also fit – the card will be 32PB of which 21 will be the content plus the index.

On the previous assumption that the size of the page plus all the components (images, video, etc) are 14 times the size of the HTML page, then by 2043 it will be possible to store the entire indexable web on a single SD card.

Section 4 examined the growth in hard disk and SD memory card capacity. Hard drives follow Moore's Law. SD cards follow Kryder's Law. Both grow faster than the indexable web and consequently it will eventually be possible to store the index (and, indeed, the indexable web itself) on a single drive. The first time this will be possible is about 2030 and it will be on a hard disk drive.

## 5. Error Estimate

Section 3 examined the current size of the Internet in indexable web pages and the size of those pages. It used a model for human population growth from the United Nations and an estimate of Internet saturation at 2050. By 2050 there will be about half a trillion web pages indexed by the search engines, accounting for 37PB of text. The inverted index will be about 11% of the size of the collection.

Storage capacity growth was examined in Section 4. Capacity will eventually exceed indexable web size. Within a decade it will be possible to index all the indexable web pages on a single hard drive. Within 2 decades it will be possible to store the documents and the index. Within three decades it will be possible to fit the index on a smart phone. Within four decades the entire static web will fit on the phone. Before 2050 the entire indexable web including images will fit. These predictions rise two questions. The first is the error in the estimates. The second is the consequences of the estimate. This section addresses the error while later sections examine the consequences.

Many of the population based estimates in Section 3 are drawn from long historical samples. For example the world population estimates date back to 1950. Projecting a historic pattern of 70 years forward by 10 is not unreasonable.

Other predictions are, however, less reliable. Internet adoption rates can only date back to the wide-scale adoption of the Internet, *circa* 1990. Projecting a 20 year history into a 10 year future is less reliable. It is reasonable, for example, to expect a steep and sudden increase in Internet use as Internet-enabled smart phones become widely adopted. Addressing this issue the extrapolation herein is already an s-shaped cubic – assuming a steep increase at present and a slower increase later. The projection also assumes Internet use will saturate at total world population which is clearly not possible – the infant, the infirm, and the illiterate cannot be considered to be document creators.

There is room for substantial error in the estimate of the number of indexable web pages. On the one hand Google claim to crawl over one trillion unique URLs, but on the other hand their index appears to have peaked at 40 billion web pages (4% of that trillion).

Hard drive projections date back 30 years to 1980, and the capacity growth has remained stable over that time. The Hitachi Ultrastar 7K3000 3 TB drive has a storage capacity of up to 3 TB with a density of 360-370 Gigabits per square inch[33]. Research into drive capacity suggests that densities up to 10 Tb/in$^2$ may be possible with conventional recoding technology (Wood, Williams et al. 2009) and that capacities of at least 100 Tb/in$^2$ are possible (Stipe, Strand et al. 2010). That is, current research suggests that a 3.5 inch drive with 5 platters could hold as much as 600TB. The 733TB index (2% of collection size) predicted for 2050 suggested in section 3.7 would not-quite-fit on a foreseeable drive, but the 4PB index (at 11%) would fit on 8 such drives.

It is reasonable to expect growth in SD cards will slow to Moore's law – and this slow down is not included in the estimates herein. The limit on Moore's law is a current topic of debate. In 2005 Moore stated "In terms of size [of transistor] you can see that we're approaching the size of atoms which is a fundamental barrier, but it'll be two or three generations before we get that far - but that's as far out as we've ever been able to see. We have another 10 to 20 years before we reach a fundamental limit. By then they'll be able to make bigger chips and have transistor budgets in the billions."[34]. In 2008 Gelsinger (an Intel Senior Vice President) stated that the law will hold until at least 2029[35]. But Moore's law only applies to the density of transistors on a silicon wafer – not the storage capacity of a device using here-to unforeseen technology and, as Moore points out, we have never been able to see more than two or three generations into the future.

If the storage capacity of an SD card continues to follow Kryder's law, but hits a fundamental limit in 2029, then the maximum storage capacity will be 32TB and 128 such cards would be needed to store the 4 PB index. That is, a grid of 11 by 12 such cards would be sufficient – or more likely a single 3D chip with transistors stacked on top of each other would be used. It is notable that existing flash RAM is based on NAND technology and that replacement technologies have already been proposed (Kim and Koh 2004).

---

[33] See the Hitachi datasheet: Ultrastar$^{TM}$ 7K3000 3.5-Inch Enterprise Hard Disk Drives
[34] http://news.techworld.com/operating-systems/3477/moores-law-is-dead-says-gordon-moore/
[35] http://java.sys-con.com/node/557154

Errors in the estimates of hard disk drive and SD card capacities do not appear to be fundamentally problematic to the projections herein, but technology changes are needed and expected.

Population growth and Internet uptake also do not appear to be a source of substantial error. The most likely source of error is the number of pages indexed.

A small total error in the estimates would not substantially alter the projected dates. Accepting an error as high as one order of magnitude (1,000%), would make it necessary for the user to wait for the hard disk capacity to increase 10 fold which, doubling every 18 months, is less than 7 years. During that time the web will have become larger, and so they must wait longer, and so on.

In Figure 17 the size of the indexable web-pages (including images, etc.) and the index, each being 10 times the estimates hereinbefore, is plotted along with estimated hard drive capacity estimated herein. If there were an error of 1,000% then by 2025 the index will fit on a single hard drive and by 2032 then text along with the index will fit, and by 2040 the whole indexable web and the index will fit. Even a 1,000% error does not substantially change the outcome of the estimates.

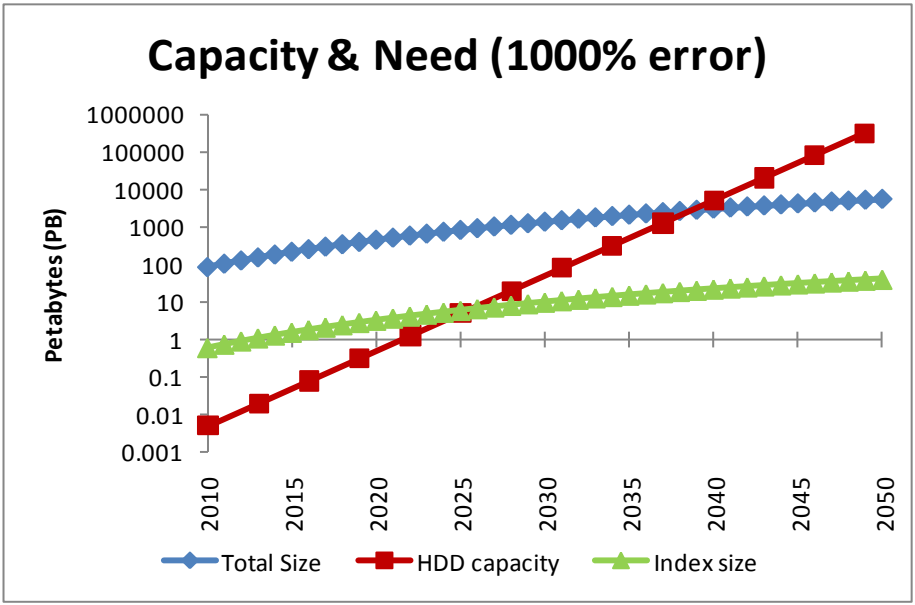

**Figure 17: Assuming a 10 fold error in the size of the web moves the date at which the index (triangles) will fit to 2025 and the date at which the full content including images (diamonds) will fit on a hard disk (squares) to 2040**

Section 3.3 examines the number of searches per month as reported by comScore. An error in these numbers affects the amount of computing power a centralized search engine needs in its data centre. If the estimates are low then the existing data centres are larger than expected. If the estimates are high the task is easier than expected and the data centres are smaller than expected. Either way it does not affect the number of client devices attached to the Internet – they are attached whether or not the user is currently searching. Either way, a decentralized search model that includes the client device as part of the search task requires a smaller data centre than a centralized model.

This section examined possible source of error in this work. It is suggested that the error is likely to be small, but nonetheless an error of 1,000% was examined. Even with such an error the outcome is not substantially affected.

## 6. Growth in CPU Speed

The increase in CPU speed between the invention of the microprocessor and the wide uptake of multi-core CPUs is well understood to have followed Moore's law. CPUs eventually hit the power wall in which exponentially more power was needed for a factorial increase in frequency. This power is given off as heat and it was not possible to remove it from the CPU. Without heat dissipation the CPU became so hot that the heat interfered with the working of the transistors. To increase the number of transistors in the same area of silicon die CPU manufacturers decreased the clock speed and opted to increase the core count thus introducing modern parallel CPUs.

The first multi-core x86 CPU from Intel was the Pentium-D which was released in May 2005 and contained 2 CPUs each on a separate chip, but housed in the same package. In March 2010 Intel released the Core-i7 980 Extreme with 6 cores, and in April 2011 the 10 core Intel Xeon E7 series (e.g. Xeon E7-8870). In October 2011 AMD released the 8-core FX-8150. Oracle similarly released an 8-core SPARC CPU in 2011. In 6 years the CPU core count has doubled twice, the equivalent to doubling every 3 years (half Moore's law). But, two sample points with core counts of fewer than 10 is hardly sufficient to draw a trend. For a vision of the future of CPUs it is necessary to look at current research rather than current market trends.

Cooling systems in desktop and smart phones are typically air-based (i.e. a fan or convection). Water based cooling systems for desktop PC have been introduced and they work much like an automobile cooling system. Heat is transferred to water from the CPU, channelled to a radiator which is cooled with a fan.

Sabry *et al.* (Sabry, Sridhar et al. 2011) recently introduced a liquid-cooled system on a chip (SoC) in which a liquid coolant was used between silicon tiers inside a 3D chip. The chip consisted of 4 layers with CPUs (each an 8-core SPARC T1) and memory (the cache) on separate layers connected by through vias. Ruch *et al.* (Ruch, Brunschwiler et al. 2011) suggest that if that liquid is an electrochemical power supply then the same channels could be used to both power the CPU (as a redox flow battery) and cool it at the same time. A prototype chip is expected in 2014, and "biological efficiencies for information processing" are expected by 2060!

IBM has already built the Aquasar water-cooled computer that has a peak performance of 10 teraflops. Michel[36], recently discussing electrochemical powered and cooled computers commented that "We currently have built this Aquasar system that's one rack full of processors. We plan that 10 to 15 years from now, we can collapse such a system in to one sugar cube - we're going to have a supercomputer in a sugar cube". If such systems materialise then the 2D Moore's law miniaturisation wall will not prove problematic to cross.

Communication between cores in a high core count multi-core CPU is also being investigated. Intel has demonstrated the Single Chip Cloud (SCC), a 48-core 32-bit Pentium CPU (Howard, Dighe et al. 2011). The chip is a distributed network of IA-32 cores connected by routers. The architecture is based on the network infrastructure seen in large data warehouses. Indeed, it is reasonable to expect that a portable device of the future to be more powerful than a data centre of today.

---

[36] http://www.bbc.co.uk/news/technology-11734909

However, the computational power of CPUs of the future does remain an area of debate. In November 2011 Vardi's editorial in CACM (Vardi 2011) suggested that current research suggests that there is an energy wall preventing further miniaturization, but did not go so far as to predict it cannot be surpassed (in fact suggesting grapheme and plasmonics might replace CMOS). In the same month Graham-Rowe (Graham-Rowe 2011), publishing in the New Scientist, suggested that a computer as powerful as IBM's Watson ("the size of 10 large refrigerators" and requiring "85 kilowatts to run") "could one day be squeezed onto mobile devices small enough to fit in your pocket".

**Table 3: LAN standards and bandwidths suggesting a 10x bandwidth increase every 5 years**

| Date | Bytes/sec | Standard | Protocol |
|------|-----------|----------|----------|
| 2010 | 12.5 GB/s | 802.3ba | 100GBASE-R |
| 2003 | 1.25 GB/s | 802.3ae | 10GBASE-SR |
| 1999 | 125 MB/s | 802.3ab | 1000BASE-T |
| 1995 | 12.5 MB/s | 802.3u | 100BASE-TX |
| 1990 | 1.25 MB/s | 802.3i | 10BASE-T |

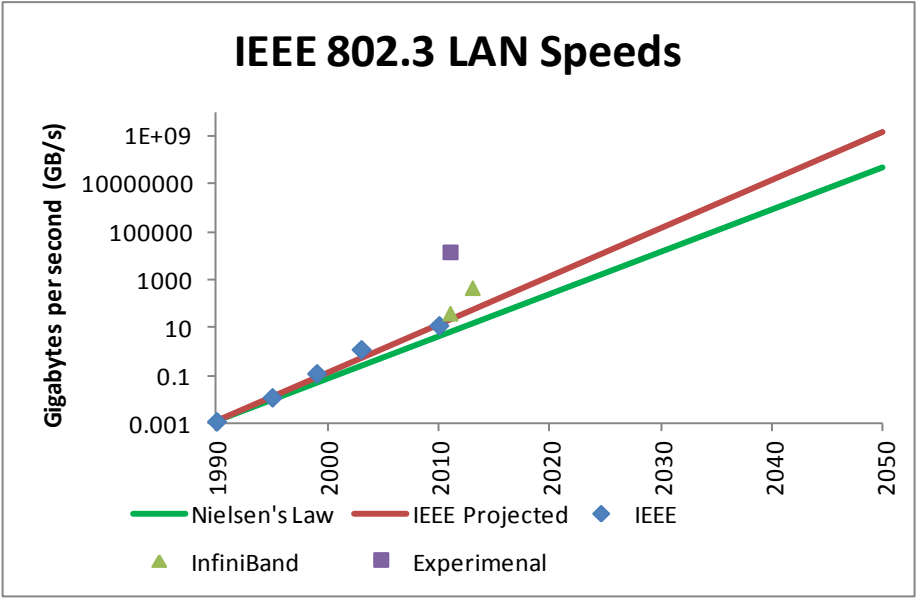

**Figure 18: LAN bandwidths projected to 2050 alongside recent experimental results**

## 7. Growth in Network Bandwidth

The growth in network (LAN) bandwidth can be plotted directly from the IEEE 802.3 standards. A list of some of those standards is given in Table 3 along with the ratification date. Figure 18 plots the date of ratification against the bandwidth along with the trend line. The trend is expected to follow Nielsen's Law that states that "high-end user's connection speed grows by 50% per year"[37], a law that Nielsen has shown to hold true since he purchased his first 300 bps modem in 1984.

LAN speed appears to grow faster than Nielsen's law predicts, increasing tenfold every five years. If this trend continues then by 2050 LAN bandwidth will be in the order of 1 Exabyte per second. By Nielsen's Law it would be only 42 Pb/s (5 Petabytes per second).

By contrast, the existing 12 channel InfiniBand EDR runs 12 parallel channels at 26 Gb/s per channel with a resulting transfer rate of about 312 Gb/s (39 GB/s). The InfiniBand Trade Association predicts data transfer rates as high as 300 Gb/s (37.5GB/s) per channel by 2013, resulting in an effective data transfer rate of 450 GB/s with 12 channels[38]. The current and projected InifiniBand transfer rates are also plotted in Figure 18 - near half-Terabyte per second data transfer rates between computers are expected with 2 years.

But faster transfer rates have already been demonstrated. Sakaguchi et al. (Sakaguchi, Awaji et al. 2011) demonstrate a transfer rate of 109 Tb/s over a single 7 core optical fibre, and Qian et al. (Qian, Huang et al. 2011) demonstrated 101.7 Tb/s. These data transfer rates of about 13 TB/s are also plotted in Figure 18 as Experimental.

To realise data transfer (throughout) rates as high as 1 EB/s will in all likelihood (at least initially) require both further research into fibre optics and parallel channel data transfer. It is unlikely that much can be done to affect latency as this is governed by the speed of light and not frequency – it will always take approximately a twentieth of a second for light to travel the approximately 13,000 mile trip half way around the world (the longest shortest path from any point to any other point on earth).

If network bandwidth continues to follow the current trend then at some point in the mid 2030s it will be possible to download the web index in less than a second. Figure 19 shows the projected growth of the web index size, the size of the web pages (including images) in Petabytes and the projected data transfer rates in PB/s. If IEEE standards follow the current trend than at by 2050 it will not only be possible to download the index sub-second, but it will also be possible to download the content too. If Nielsen's law is followed then it will possible to download an index sub-second but not to download the content sub-second.

It is important to make a distinction between all possible machine-generated content and the indexable-web. The indexable web that is included in the discussion herein only includes those pages that the most popular search engine currently index – and their attachments. It does not include automatically generated content that comes from such devices as the Australian Square Kilometre Array Pathfinder which is expected to be processing 72Tb/s feeds coming from 36 antennas and producing 2.8GB/s of processed data for storing (producing 1PB of data to be stored every 4 days) (Cornwell and Humphreys 2010).

## 8. New Search Architectures

Hereinbefore it has been shown that the indexable web and its index will fit on a single storage device in the near future. This section discusses new search engine architectures that take advantage of this observation.

---

[37] http://www.useit.com/alertbox/980405.html
[38] http://www.infinibandta.org/content/pages.php?pg=technology_overview

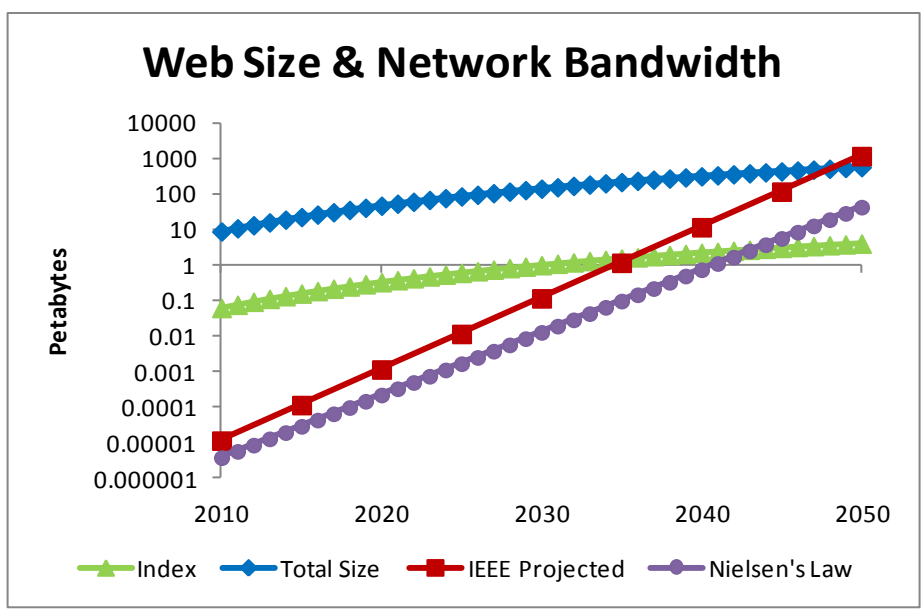

**Figure 19: Projected network bandwidth in PB/s by IEEE standards (squares) and Nielsen's law (circles) suggests that by 2050 it will be possible to download the index of the web (triangles) in less than a second and perhaps even the entire content of the indexable web including images (diamonds)**

Some hard disk drives and memory cards are already shipped with content. A user purchasing a new PC expects the machine to boot from the hard drive without having to first install the operating system from other media. The operating system is pre-loaded on the drive.

Hard disk drive manufacturers could pre-install not only the operating system but also an index for a search engine as well as documents and the search engine. The same is true for smart phone manufacturers – it is not unusual for a phone's memory card to come pre-loaded with demo versions of software. Adding a collection of web pages, images, an index, and a search engine is not infeasible.

With such a device (a pre-loaded computer or smart phone) the user could search a snapshot of the web regardless of that user's Internet connectivity. This is of immediate utility. For example, lives could be saved if all previously published medical literature were pre-loaded onto smart phones given to medics in hard to reach, or non-Internet connected locations. The ability to carry all of science in the pocket would make research possible to mobile users on a plane or train. A stove might be shipped with a display and interface capable of searching all known recipes. A smart phone might be used to identify toxic or safe wild-plants. Indeed, the utility ranges from general household tools to specialized devices for industrial or medical use.

Missing from this design is the ability to keep the system up-to-date, however this is not problematic. Prior research (Charzinski 2010) has shown that client side caching can reduce the total number of elements downloaded by 75% and the total number of bytes by 81%. On average, more than 50% of downloaded content is served locally when client side caching is used. Web designers are encouraged to make their web sites cacheable, and many do. Web content is cached not only by the client computer but also by proxies held at many locations between the client and the server (including the user's ISP). Most users are already seeing a substantial proportion of the web as it was, not as it is!

To remain up to date a user could periodically download a fresh Internet dump (and index). But in the case of a smart phone it is most reasonable to expect the user to perform this by periodically discarding their phone and purchasing another. If the current pattern is maintained, this will be approximately every 18 months (on average). That is, on average the user's copy of the Internet will be half that in age, or only 9 months old. For all but searching on current events, this could be sufficient for most users. Despite this, new distributed search architectures capable of remaining up-to-date are proposed.

Current search engines are designed to cater to millions of users searching billions of pages. To do this millions of computers are connected together with high speed communications cables, and globally distributed data centres are used. The scalability of this model has already been questioned by Baeza-Yates (Baeza-Yates 2010).

An alternative solution is to off-load some of the search load to the user. This might be done by pre-loading most of the indexable web (certainly those pages with a very low rate of change) to the user's device and only online-searching the pages with a high rate of change.

This distributed search engine would integrate the user's client computer with a remote data centre. The client software would transmit a snapshot date to the data centre which would then search only through the differences (between the snapshot date and the current date) and transmit the results back to the user. Such snapshots might be taken every month, or even every day. The snapshot date could be indirectly encoded – for example it might be a checksum of the software, or it might be the serial number of the client's computer (or CPU). Or it might be the MAC address of the network card. These could be converted into a date through table lookup.

The detailed architecture of the data centre is left for future work; however one such architecture takes advantage of existing distributed information retrieval principles.

At present many search engine companies divide the document collection into a number of approximately equal sized (in documents) chunks called shards. Each shard is then allocated to a separate computer. When a query arrives at the data centre it is distributed to shards by a broker and the results from each broker are merged before finally being sent back to the user. There are often many replicas of the shard-set in a single data centre, each concurrently serving many clients.

If documents are assigned to shards in a uniform but random way then the per-query load is naturally balanced because the postings lists on each shard are about the same length. If a single shard fails for such reasons as a hardware fault then the remaining shards can be expected to be approximately as good as the whole set because relevant documents are near-uniformly distributed across the shards. If the search load peaks then load balancing can be performed by simply reducing the number of shards used to resolve the query.

Assuming the user has a fixed-dated snapshot of the indexable web, and assuming documents are distributed to shards on a modification-date basis then all the documents representing the difference between the snapshot and the current date would be confined to a known subset of the shards. There would be no need to search older shards.

To support two users with different snapshots of the web, each would transmit their snapshot date to the data centre which would compute the shards necessary for each. They need not, and indeed would not, typically be the same subset of shards.

The proposed architecture is illustrated in Figure 20 (for the year 2026). The data in the data centre is divided into year-sized shards dated 2021-2026. Each shard set is replicated four times. There are two users, *A* and *B*. *A* has a snapshot on their smart phone dated 2025 and only needs to search on their local device and the 2026 differences. User *B* has a snapshot dated 2023 and so needs to search those documents from 2024 – 2026.

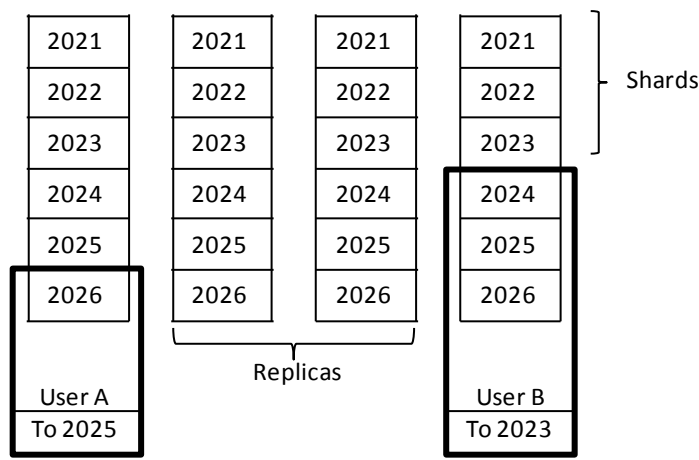

**Figure 20: Users A and B have differently dated snapshots and so search different sets of shards in the data centre**

As users update their devices the load on older shards in the data centre would be reduced. A consequence of this is that the shards could be replicated asymmetrically. That is, more replicas of newer shards and fewer replicas of older shards. Eventually all devices using the extremely old shards would become obsolete and the content on those shards could be deleted from the data centre. The data centre would no longer be a forever-growing archive of the sum of the web but would be a sliding window over time.

This architecture provides a fully distributed search engine that can be used either online or offline. When used online it will be up-to-date. Either way it will reduce the load on the data centre by offloading the search to the user's device. As such it is inherently a more scalable architecture than the current centralized data centre designs. The client's device is already an integral part of the search process because the user enters the query on it. This model proposes to use it to also perform most of the search work thus reducing the load on the data centre.

### 8.1. CPU Requirements

Neither Google nor Bing release data on the number of shards (computers) used to perform each query. Dean (Dean 2009) suggests that there are thousands. Section 6 discussed past and future trends in CPU computational power – IBM are suggesting that in the future a CPU as powerful as a modern supercomputer will be no larger than "a sugar cube", and Intel are experimenting with 48-core CPUs already. A 1000-core general purpose CPU is not beyond the imaginable, indeed Adapteva[39] have announce their design for a 1000-core general purpose CPU, and Tilera[40] have a 100-core general purpose CPU already on the market. As text search is an embarrassingly parallel problem, search latency can be expected to decrease as more cores with increasingly large caches are added to the CPU. Indeed, with several thousand general purpose cores on a single CPU the round trip time to the data centre may take longer than searching the archive on the smart phone. In this case, sending a message to a data centre every time the user performs a search could become the bottleneck thus making it more attractive to upload the index directly to the client for local search there.

---

[39] http://www.adapteva.com/
[40] http://www.tilera.com/

## *8.2. Network Requirements*

With a network bandwidth of 1 EB/s it would take less than a second to download a 4 PB index of the 2050 web to a smart phone. This raises the question of whether or not it is worthwhile having the data centre at all. A user might, for example, simply connect their device to the wall to re-charge the batteries and download any and all updates to their personal archive at the same time (assuming wired power remains). If the update is sub-second then it might take less time to sync than to charge the battery. The current shift to cloud computing and the storing (archiving) all files in large centralized data centres could make this easier as each cloud could keep a transaction log of all changes made on it and ship that to the user (rather than the contents of *all* files) on request. In this case the hand-held device would act as a mirror of all data on the Internet rather than simply a thin-client to it. This model is illustrated Figure 21 where, for example, User A connects to cloud C1 to transmit their transaction log and to receive a log of all changes made by all other users on all other clouds (C1, C2, C3).

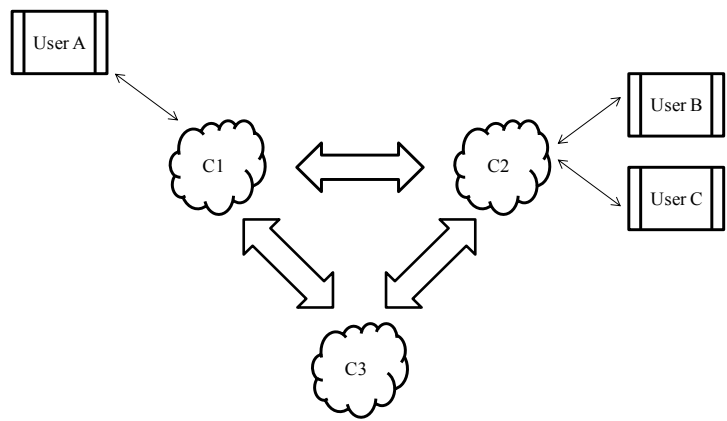

**Figure 21: Users connect to clouds to receive transaction logs**

If all users were constantly connected to each and every other user at all times (much as a mobile phone is constantly connected to the cell network) then with very high data transfer rates it might be possible to transmit to all users all changes made by all other users to all documents (access control permitting). In this way the Internet would resemble a star network. Each device would be responsible for correctly maintaining all files owned by all users and keeping the indexes up to date. If this were the case then all Internet messages could be *broadcast* to all other users rather than being sent point to point (reducing the bandwidth requirements) – however if a client device missed a message for some reason then it might be difficult to "go back and get it" unless it were stored somewhere centrally. In this model computers in the data centre would be there to archive all changes made by all people on all devices (the historical value of which should not be underestimated). This third model is illustrated in Figure 22.

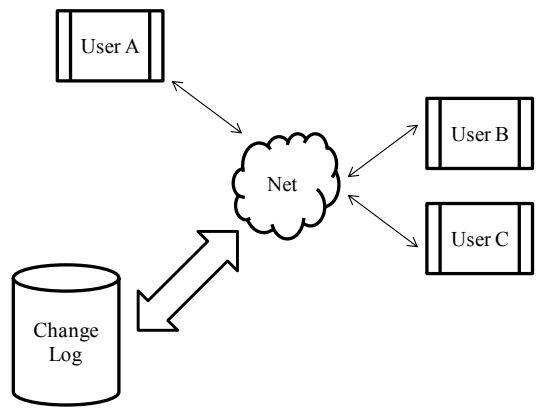

**Figure 22: Users broadcast all changes to each other, but those changes are archived so that any lost messages to be received.**

## *8.3. Discussion*

Having the entire indexable web stored and searchable on a smart-phone could fundamentally change the way search results are ranked. Present ranking strategies take user click patterns into account and rank based on group clicking behaviour – a click to see a document is a vote for its relevance. If each user had their own instance of the index then such a ranking strategy would have to be pre-computed based on data sourced from each user's local search patterns.

The user would gain a certain amount of security and privacy by refusing to transmit that information back to the service provider. For the general public this is likely to be of little concern, but it is for in the corporate environment where there is a certain amount of paranoia that someone snooping search requests could steal industrial secrets. Such an industry is the drug discovery and patent industry.

The funding model for Internet search is currently based on targeted advertising. It is not clear how search would be funded in the proposed model. An advertising model could be built into the software on the smart-phone, or alternatively the update service could be a subscription service. But, as the life expectancy of a smart-phone is only short, the cost could also be built into the purchase price of the phone. Removing advertising from search would also affect the advertisers, but it is beyond the scope of this contribution to speculate on the state of the economy 40 years from today.

The proposed model for web search is based on an assumption that web search will be very much the same in 2050 as it is today – an unverifiable hypothesis. Today we see streams of data such as Facebook, Google+ and Twitter being adopted as standard forms of human to human communication. This kind of data is not excluded from the model herein. Although indexable web pages have been used for the analysis, a tweet or a social media post is typically a string of text and as far as the search engine is concerned, another indexable object that when clicked sends the user to an external site to retrieve a page. However, it is reasonable to assume a search engine implementer will choose to deal with these objects in a different way to static web pages – much as search engines that index the Usenet currently treat that data differently from web pages. Indeed, it is not inconceivable to believe that the search engine of 2050 will be more attune to dynamically changing data. Research on searching social media is ongoing by others. Research on dynamic indexes has been conducted for many years and several strategies have already been proposed (Lester, Zobel et al. 2004).

In Section 8.2 (and in Figure 22) it is suggested that all changes to all documents could be broadcast to all users, and that each user has an up-to-date index as a consequence (private messages could be pushed to just the recipients). Once network bandwidth exceeds the human capacity to generate new information this model becomes appealing. Our smart phones will constantly listen for new information and constantly keep their index up-to-date.

However, it seems unreasonable to believe that static web pages (or the equivalent) will disappear by 2050. At present the Wikipedia is one of the most heavily visited sites and it consists of a set of several million (essentially) static pages. Despite the enormous quantities of text generated through Twitter and Facebook, the human demand for encyclopaedic information persists. Encyclopaedia Britannica was first published in 1768 (Chambers's Cyclopaedia in 1728), and the demand for this kind for information has not disappeared in those nearly 300 years – another 38 years (13%) is unlikely to see demand disappearing. Although methods of dialog have changed, and continue to change, the scientific archive also continues to be an archive.

## 9. Conclusions and Future Work

In order to build a search engine it is first necessary to place bounds on the problem. Herein the expected size of the web is computed by examining estimates of the population growth, Internet uptake, web page size, and web page counts.

It is shown that sometime in the next decade it will be possible to store the index of the entire indexable web on a single desktop computer's hard drive. Within another decade it will be possible to store the text, and within another the whole indexable web and its index. Not long after, it will be possible to store all this on a mobile phone flash memory card.

New search architectures that include the computing power of the user's device are proposed. These architectures involve preloading a snapshot of the indexable web onto the user's device and searching the snapshot there. In the case that the user is off-line the results would be old. In the case where the user is on-line the results would be supplemented (and amended) by additional results from a web search engine.

Changes to the sharding algorithm of the data centre are proposed. Rather than distributing documents to shards on a random but uniform way, they would be distributed to shards by date. Computing resources would then be assigned to shards in a load-based way with most resources being applied to the most recent content. As users discard old devices the need to search old shards would be reduced and eventually old shards would become redundant.

If all devices were permanently connected then they could be kept up-to date in the background. In this case there is no need for a data centre other than to archive messages.

A consequence of offloading search to the user's device is that the data centres necessary for web search will be reduced in size (or even become unnecessary). This reduction is also a reduction in cost, and in centralized energy consumption. Not only will users be able to search the Internet when offline but when online the cost will also be substantially reduced.